\documentclass{article}

\usepackage{arxiv}

\usepackage[utf8]{inputenc} 
\usepackage[T1]{fontenc}    
\usepackage{hyperref}       
\usepackage{url}            
\usepackage{booktabs}       
\usepackage{amsfonts}       
\usepackage{nicefrac}       
\usepackage{microtype}      
\usepackage{lipsum}
\usepackage{graphicx}
\graphicspath{ {./images/} }
\usepackage{siunitx}
\usepackage[utf8]{inputenc}
\usepackage{placeins}
\usepackage{times}
\usepackage{multirow}
\usepackage{amsmath}
\usepackage{graphicx}
\usepackage{tabularx}
\usepackage{lineno}
\usepackage[export]{adjustbox}
\usepackage{amsmath}
\usepackage{epstopdf}
\usepackage{xcolor}
\definecolor{amber}{rgb}{1.0, 0.75, 0.0}
\definecolor{brickred}{rgb}{0.7960, 0.2550, 0.3290}
\usepackage{placeins}
\usepackage{times}
\usepackage{multirow}
\usepackage{amsmath,amssymb}

\newcommand{\subfigimgtwo}[3][,]{%
  \setbox1=\hbox{\includegraphics[#1]{#3}}
  \leavevmode\rlap{\usebox1}
  \rlap{\hspace*{-5pt}\raisebox{\dimexpr\ht1-1\baselineskip}{#2}}
  \phantom{\usebox1}
}
\newcommand{\subfigimgthree}[3][,]{%
  \setbox1=\hbox{\includegraphics[#1]{#3}}
  \leavevmode\rlap{\usebox1}
  \rlap{\hspace*{-10pt}\raisebox{\dimexpr\ht1-1\baselineskip}{#2}}
  \phantom{\usebox1}
}

\usepackage{tikz}

\usepackage{subcaption}
\usepackage{mathtools}
\usepackage{bm}
\usepackage[utf8]{inputenc}
\usepackage{booktabs}

\newcommand{\citet}[1]{\citeuthor{#1} \shortcite{#1}}

\title{Parameter uncertainty quantification in an idealized GCM with a seasonal cycle}

\author{
 Michael F. Howland \\
  Civil and Environmental Engineering \\ Massachusetts Institute of Technology \\ Cambridge, MA, USA \\
  \texttt{mhowland@mit.edu} \\
   \And
   Oliver R. A. Dunbar \\
  Division of Geological and Planetary Sciences \\
  California Institute of Technology \\ Pasadena, CA, USA \\
  \And
  Tapio Schneider \\ 
  Division of Geological and Planetary Sciences \\
  California Institute of Technology \\ Pasadena, CA, USA \\
}

\begin{document}
\maketitle
\begin{abstract}
Climate models are generally calibrated manually by comparing selected climate statistics, such as the global top-of-atmosphere energy balance, to observations.
The manual tuning only targets a limited subset of observational data and parameters.
Bayesian calibration can estimate climate model parameters and their uncertainty using a larger fraction of the available data and automatically exploring the parameter space more broadly.
In Bayesian learning, it is natural to exploit the seasonal cycle, which has large amplitude, compared with anthropogenic climate change, in many climate statistics. In this study, we develop methods for the calibration and uncertainty quantification (UQ) of model parameters exploiting the seasonal cycle, and we demonstrate a proof-of-concept with an idealized general circulation model (GCM). Uncertainty quantification is performed using the calibrate-emulate-sample approach, which combines stochastic optimization and machine learning emulation to speed up Bayesian learning.
The methods are demonstrated in a perfect-model setting through the calibration and UQ of a convective parameterization in an idealized GCM with a seasonal cycle. 
Calibration and UQ based on seasonally averaged climate statistics, compared to annually averaged, reduces the calibration error by up to an order of magnitude and narrows the spread of posterior distributions by factors between two and five, depending on the variables used for UQ.
The reduction in the size of the parameter posterior distributions leads to a reduction in the uncertainty of climate model predictions. 
\end{abstract}


\section{Introduction}

The objective of quantifying uncertainty in computational models arises in a wide range of applications, including weather and climate modeling \cite{schneider2017earth}, fluid dynamics \cite{duraisamy2019turbulence}, and energy systems \cite{constantinescu2010computational}.
Often, uncertainty associated with predictions from computational models is the result of processes that cannot be resolved on the computational grid, either due to computational complexity limitations \cite{meneveau2000scale} or due to uncertainty associated with the process itself \cite{schneider2017climate}.
In general circulation models (GCMs), primary uncertainties arise from the representation of subgrid-scale turbulence, convection, and cloud physics, which have a significant impact on the evolution of climate under rising greenhouse gases \cite{cess1989interpretation, bony2005marine, webb2013origins, suzuki2013evaluating, brient2016constraints}.
While clouds are associated with turbulence and convective updrafts with scales of $\mathcal{O}(10 \ \mathrm{m})$, modern climate simulations have a typical horizontal resolution of $\mathcal{O}(10 \ \mathrm{km})$--$\mathcal{O}(100 \ \mathrm{km})$ \cite{delworth2012simulated, ipcc1p5degreeSI3}.
Climate simulations rely on physically motivated parameterizations that model the effects of subgrid-scale processes such as clouds and turbulence on the resolved scales \cite{hourdin2013lmdz5b}. Such parameterizations come with parametric and structural uncertainties; quantifying these uncertainties and how they percolate into climate projections remains an outstanding challenge \cite{schneider2017climate}.

Physical parameterizations have historically been individually developed and calibrated using data from isolated experiments \cite{smagorinsky1963general, spalart1992one, Jakob10a, golaz2013cloud,  hourdin2017art}. They are further adjusted so that global models that incorporate them satisfy selected large-scale  observational or physical constraints, such as a closed top-of-atmosphere energy balance or reproduction of the 20th-century global-mean temperature evolution \cite{Mauritsen12a,hourdin2017art,schmidt2017practice}. 
Model calibration is usually done manually, focusing on a small subset of model parameters and exploiting only a fraction of the available observational data. 

As a step toward automating and augmenting this process, here we further develop algorithms for model calibration and uncertainty quantification (UQ) that in principle allow models to learn from large datasets and that scale to high-dimensional parameter spaces. In previous work, these algorithms have been demonstrated in simple conceptual models \cite{cleary2021calibrate} and in a statistically stationary idealized GCM \cite{dunbar2020calibration}. 
We take the next step and demonstrate how these algorithms can exploit seasonal variations, which for many climate statistics are large relative to the climate changes expected in the coming decades and contain exploitable information about the response of the climate system to perturbations \cite{knutti2006constraining,Schneider21b}. 

Whereas numerical weather prediction assimilates spatiotemporally evolving trajectories of atmospheric states as initial conditions for forecasts (e.g., \cite{Kalnay03a, bauer2015quiet}), in climate modeling it is preferable to assimilate time-averaged climate statistics. This focuses the learning problem on quantities of interest in climate predictions (i.e., climate statistics, including higher-order statistics such as precipitation extremes), and it avoids the need to estimate uncertain atmospheric initial conditions on which trajectories of states depend \cite{cleary2021calibrate}. Calibration and UQ of climate models on the basis of time-averaged statistics smooths the prediction-error based objective function and enables the use of data that have different resolution than the climate simulations \cite{schneider2017earth, dunbar2020calibration}. Given the dual desires to avoid having to estimate atmospheric initial conditions, which are forgotten over about 2 weeks \cite{BuiCheEmaMagSunZha19}, and to exploit seasonal variations, it becomes natural to choose averaging timescales between around 30 and 90 days. Such averaging timescales are the focus of this study. 

We consider the calibration and UQ of convective parameters in an idealized GCM with seasonally varying insolation \cite{frierson2006gray,o2008hydrological,bordoni2008monsoons}.
We develop an extension of the calibrate-emulate-sample (CES) Bayesian learning methodology \cite{cleary2021calibrate} to enable the use of statistics computed from a non-stationary statistical state.
The qualitative and quantitative impacts of the time-averaging length of the climate statistics on the parameter calibration and UQ are assessed.
GCMs are generally tuned in situations which have low parameter identifiability based on available climate data \cite{hourdin2017art}. We perform
numerical experiments with observable climate statistics that are informative about the convective parameters we wish to calibrate, but are not in any simple and direct way related to them. 
We also explore less informative climate statistics, which highlight the benefits of incorporating seasonal variations in the climate statistics that are being exploited for UQ.

The remainder of this paper is organized as follows.
In section~\ref{sec:ces}, the Bayesian learning methods for time-dependent problems are introduced.
The numerical details of the seasonally forced GCM and UQ experiments are introduced in section~\ref{sec:gcm}.
The calibration and UQ results are shown in section~\ref{sec:results}, and conclusions are provided in section~\ref{sec:conclusions}.

\section{Uncertainty quantification methods}
\label{sec:ces}

The goal of this study is to estimate the probability distributions associated with model parameters $\bm{\theta}$ that are used by a GCM and about which only imprecise prior information is known.
We will consider a Bayesian approach to the estimation of the probability distributions of model parameters $\bm{\theta}$, where we seek $\mathbb{P}(\bm{\theta} | \bm{y})$, the conditional probability distribution of parameters $\bm{\theta}$ given observed data $\bm{y}$.
The GCM is a computationally expensive numerical model that evolves climate states in time.
From the GCM, we extract statistical information that is denoted by $\mathcal{G}(\bm{\theta})$.
Here, $\mathcal{G}(\bm{\theta})$ includes the numerical integration of the GCM states and the aggregation of relevant climate statistics in time.
Although the data $\bm{y}$ will, in general, not provide direct information about $\bm{\theta}$, we can estimate $\mathbb{P}(\bm{\theta} | \bm{y})$ using Bayesian learning by comparing the GCM outputs $\mathcal{G}(\bm{\theta})$ with data $\bm{y}$.

\begin{figure}
    \centering
    \includegraphics[width=0.9\linewidth]{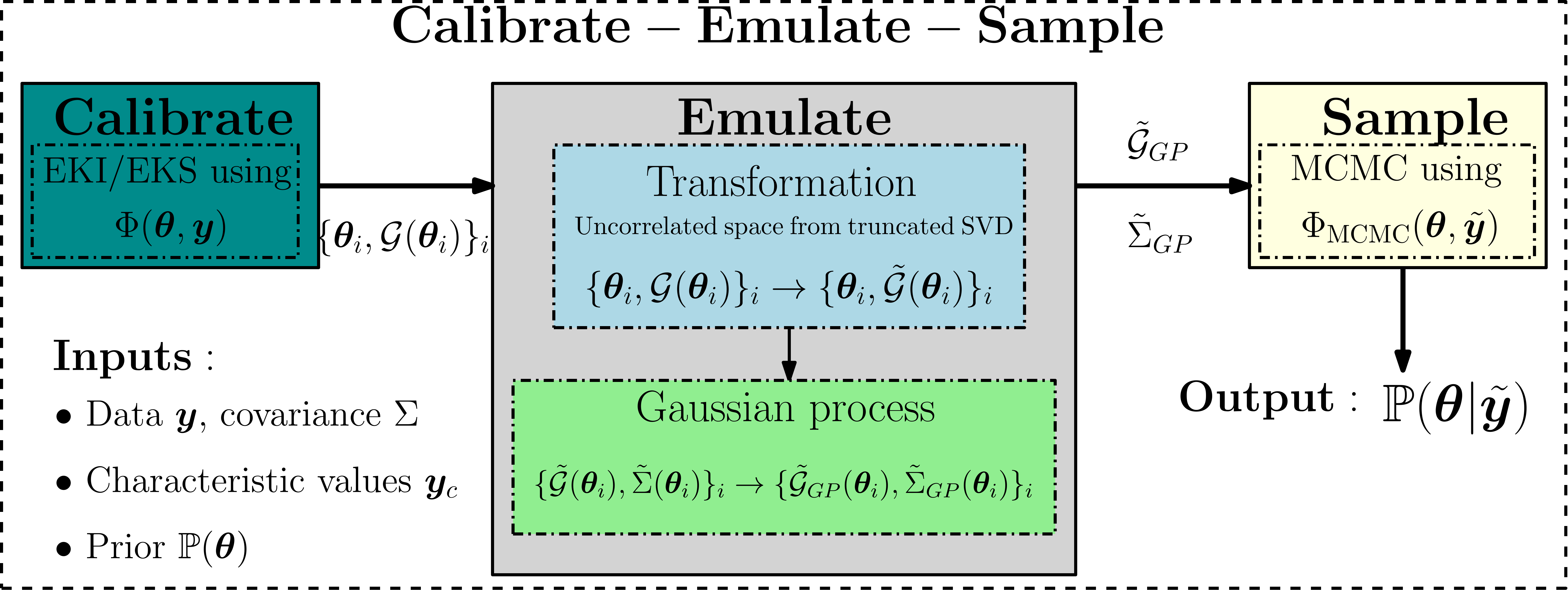}
    \caption{
    Schematic of the calibrate-emulate-sample (CES) methodology to estimate model parameters $\bm{\theta}$.
    With inputs of data $\bm{y}$, noise covariance $\Sigma$, characteristic values of the data $\bm{y}_c$ (for normalization), and the prior $\mathbb{P}(\bm{\theta})$, the calibration stage generates input-output pairs $\{\bm{\theta}_i, \mathcal{G}(\bm{\theta}_i)\}_{i}$.
    The input-output mapping is emulated using Gaussian process regression in a transformed, uncorrelated space ($\tilde{\cdot}$), obtained from a truncated singular value decomposition on the noise covariance matrix $\Sigma$.
    The GP emulator is used for efficient sampling using MCMC to approximate the posterior distribution $\mathbb{P}(\bm{\theta} | \tilde{\bm{y}})$.
    The objective functions for calibration and sampling are denoted by $\Phi(\bm{\theta},\bm{y})$ and $\Phi_{\mathrm{MCMC}}(\bm{\theta},\tilde{\bm{y}})$, respectively.
    }
    \label{fig:ces}
\end{figure}

Standard methods for Bayesian learning include Markov Chain Monte Carlo (MCMC) \cite{brooks2011handbook}, which typically requires $\mathcal{O}(10^5)$ forward model evaluations of $\mathcal{G}(\bm{\theta})$ to sample the posterior distribution \cite{geyer2011introduction}.
Instead, we perform calibration and UQ using the recently developed CES methodology \cite{cleary2021calibrate}, which consists of three steps: (1) Ensemble Kalman processes \cite{schillings2017analysis, garbuno2020interacting} are used to calibrate parameters and to generate input-output pairs of the mapping $\bm{\theta} \mapsto \mathcal{G}(\bm{\theta})$; (2) Gaussian process (GP) regression is used to train an emulator $\mathcal{G}_{\mathrm{GP}}(\bm{\theta})$ of the mapping $\bm{\theta} \mapsto \mathcal{G}(\bm{\theta})$ using the training points generated in the calibration step; and (3) MCMC sampling with the computationally efficient GP emulator $\mathcal{G}_{\mathrm{GP}}(\bm{\theta})$ rather than the expensive forward model $\mathcal{G}(\bm{\theta})$ is used to estimate the posterior distribution $\mathbb{P}(\bm{\theta} | \bm{y})$.
The CES methodology has previously been used for calibration and UQ of parameters in simple model problems such as Darcy flow and Lorenz systems \cite{cleary2021calibrate} and for convective parameters in a statistically stationary GCM \cite{dunbar2020calibration}.
More recent methodological developments by \cite{lan2021scaling} enabled the CES framework to perform simultaneous UQ on $\mathcal{O}(1000)$ parameters using deep neural network-based emulation and MCMC suited to high-dimensional spaces \cite{beskos2008mcmc, beskos2011hybrid}. A schematic of the CES methodology is shown in Figure \ref{fig:ces}.

The Bayesian learning methodology used in this study is described in the following sections.
In section~\ref{sec:seasonal_inverse}, the inverse problem of estimating $\mathbb{P}(\bm{\theta} | \bm{y})$ in a setting with a periodic cycle is introduced.
In section~\ref{sec:calibrate}, the ensemble Kalman process calibration method is outlined. Section~\ref{sec:emulate} introduces the GP emulation in an uncorrelated transformed space, obtained by a  singular value decomposition (principal component analysis) of the noise covariance matrix. 
Section~\ref{sec:sample} describes how the Bayesian posterior distribution is approximated using the GP emulator.

\subsection{Seasonal GCM inverse problem}
\label{sec:seasonal_inverse}

With fixed insolation \cite{frierson2006gray, o2008hydrological}, GCM statistics are stationary and ergodic.
With the seasonal cycle incorporated, the insolation varies as a function of the ordinal day in the simulation.
The resulting GCM states are statistically cyclostationary, with a dependence on the ordinal day.
The data we use are constructed accounting for the seasonally varying boundary conditions, as the time-averaged GCM statistics
\begin{equation}
\mathcal{G}_T(\bm{\theta}, \xi, t) = \frac{1}{T} \int_t^{t+T} \mathcal{H}(t^\prime; \bm{\theta}, \xi) dt^\prime.
\label{eq:timeAvg}
\end{equation}
Here, $\mathcal{H}(t; \bm{\theta}, \xi) \in \mathbb{R}^{N_y}$ represents the output states of the GCM, depending on time $t$.
The number of states being measured is $N_y$.
The integration length is specified as $T$.
The initial conditions for the GCM integration are represented by $\xi$, which depend on the initial time $t$.
The time-averaged statistics are $\mathcal{G}_T(\bm{\theta}, \xi, t) \in \mathbb{R}^{N_y}$.

With seasonally varying insolation, the time-averaged statistics in Eq. \eqref{eq:timeAvg} depend on the ordinal day ($t$) in the GCM.
With a specified integration length $T$, there are a corresponding set of time-averaged statistics
\begin{equation}
\mathcal{G}_{T,j}(\bm{\theta}, \xi) = \frac{1}{T} \int_{t_0+(j-1)T}^{t_0+jT} \mathcal{H}(t^\prime; \bm{\theta}, \xi) dt^\prime,
\label{eq:g_season}
\end{equation}
where $j=1$ through $N_s$ is an index representing the time of year in the GCM simulation, starting from $j=1$ at vernal equinox (time $t_0$).
The integration windows are non-overlapping.
The length of the year in the GCM is $T_{yr}=360~\mathrm{d}$, and the resulting number of non-overlapping time-averaged statistics are $N_s = T_{yr} / T$.
In this framework, the number of batches of statistics $N_s$ is a design parameter set by the integration timescale $T$.
For $T=90~\mathrm{d}$, the data are aggregated seasonally, with $N_s=4$.
For $T=360~\mathrm{d}$, the statistics are averaged over the full year in the GCM, corresponding to annually averaged climate statistics, and $N_s=1$.

The GCM data are the concatenation of the time-averaged batches over collated ordinal days
\begin{equation}
\mathcal{G}(\bm{\theta},\xi) = [\mathcal{G}_{T,1}(\bm{\theta},\xi_1), \mathcal{G}_{T,2}(\bm{\theta},\xi_2), ..., \mathcal{G}_{T,N_s}(\bm{\theta},\xi_{N_s})].
\label{eq:g}
\end{equation}
In this study, we will consider perfect-model numerical experiments, with the data $\bm{y}$ being constructed using the same GCM at target true parameters $\bm{\theta}=\bm{\theta}^\dagger$.
The synthetic data are given by
\begin{equation}
\bm{y} = [\mathcal{G}_{T,1}(\bm{\theta}^\dagger, \xi_1), \mathcal{G}_{T,2}(\bm{\theta}^\dagger, \xi_2), ..., \mathcal{G}_{T,N_s}(\bm{\theta}^\dagger, \xi_{N_s})].
\end{equation}
The size of $\bm{y}$ and $\mathcal{G}$ is $\mathbb{R}^{N}$, where $N = N_s \cdot N_y$.
The resulting inverse problem relating parameters and data is
\begin{equation}
\bm{y} = \mathcal{G}(\bm{\theta},\xi) + \eta,
\label{eq:forward_ics}
\end{equation}
where $\eta \sim N(0,\Delta)$ is a realization of normal measurement error with zero mean and covariance matrix $\Delta$.
We generate synthetic data and forward model data starting from the same initial conditions $\xi$ in Eq. \eqref{eq:forward_ics}.
Since the states of the GCM depend on the ordinal day, the averaging operation in Eq. \eqref{eq:g_season} must be consistently aligned in ordinal day between the synthetic data and GCM outputs.

The states of the GCM depend on the boundary conditions, specific to the ordinal day, and the initial conditions $\xi$. 
The ensemble average of independent realizations of $\mathcal{G}(\bm{\theta}, \xi_i)$ over different initial conditions $\xi_i$ is
\begin{equation}
\mathcal{G}_\infty(\bm{\theta}) = \lim_{M\rightarrow \infty} \frac{1}{M} \sum_{i=1}^{M} \mathcal{G}(\bm{\theta},\xi_i).
\end{equation}
In the limit of infinite realizations of the climate statistics, the dependence of the ensemble average $\mathcal{G}_\infty(\bm{\theta})$ on the initial conditions trends to zero by the central limit theorem.
The central limit theorem applies since $\mathcal{G}(\bm{\theta},\xi_i)$ contains a full year of data by construction (Eq. \ref{eq:g}) and is therefore a statistically stationary object with respect to the cycle.
But due to computational limitations, only a finite number of ensemble realizations is available in practice.
For the synthetic data, we can similarly define $\bm{y}_{\infty}$; however, $\bm{y}_{\infty}$ is generally not accessible from observations. 

The atmospheric initial condition is of minimal practical value in the climate model setting, and we aim to avoid its estimation.
We therefore reformulate the inverse problem such that $\mathcal{G}$ and $\bm{y}$ can have different initial conditions, and such that we do not require the estimation of the initial conditions.
To do so, we assume the forward model output $\mathcal{G}(\bm{\theta}, \xi)$, is a noisy, finite average approximation of the infinite ensemble average of the climate statistics, that is, $\mathcal{G}(\bm{\theta}, \xi) = \mathcal{G}_\infty(\bm{\theta}) + N(0,\Sigma)$, where $\Sigma$ is the internal variability covariance matrix of the GCM \cite{cleary2021calibrate}.
The realizations $\mathcal{G}(\bm{\theta}, \xi)$ and $\bm{y}$ are both subject to the internal variability of the climate system.
In this study, given the definition of $\bm{y}$ and $\mathcal{G}(\bm{\theta},\xi)$, the internal variability is the interannual variability of the GCM.
As a result, the inverse problem becomes
\begin{equation}
\bm{y} = \mathcal{G}_\infty(\bm{\theta}) + \gamma,
\label{eq:inverse_initial_differ}
\end{equation}
where $\gamma \sim N(0,\Delta + \Sigma)$.
In Eq. \eqref{eq:inverse_initial_differ}, the initial conditions are removed from the inference problem.
The synthetic data are collected into the matrix $Y\in \mathbb{R}^{N \times N_d}$, where $N$ is the dimension of the data space and  $N_d$ is the number of data samples, or years in this study.
The internal variability covariance matrix is computed from the synthetic data matrix $Y$ such that $\Sigma = \mathrm{cov}(Y)$.
We assume that $\Sigma$ does not vary as a function $\bm{\theta}$ such that $\Sigma(\bm{\theta}^\dagger)\approx \Sigma(\bm{\theta})$ and that $\Sigma$ does not depend on initial conditions $\xi$.
Invoking Gaussian error assumptions based on the central limit theorem, the corresponding negative log-likelihood objective function then is
\begin{equation}
\Phi(\bm{\theta}, \bm{y}) = \frac{1}{2} \| \bm{y} - \mathcal{G}_\infty(\bm{\theta}) \|^2_{\Delta + \Sigma}
\end{equation}
where $\| \cdot \|_A = \| A^{-1/2} \cdot \|_2$.
The likelihood is \cite{kaipio2006statistical}
\begin{equation}
\mathbb{P}(\bm{\theta} | \bm{y}) \propto \exp(-\Phi(\bm{\theta}, \bm{y})).
\label{eq:posterior}
\end{equation}

In this perfect-model setting, where the synthetic data and forward model are obtained from the same GCM, there is no direct measurement error ($\eta$) and no systematic model error.
To emulate measurement error, we add Gaussian noise to the GCM output, with zero mean and covariance matrix
\begin{equation}
\Delta = \mathrm{diag}(\delta_k^2).
\end{equation}
Here, the noise standard deviation is defined such that 
\begin{equation}
\delta_k = \min\left(C\min\left[\mathrm{dist}(y_{k}+2\sqrt{\Sigma_{kk}},\partial \Omega_k), \mathrm{dist}(y_{k} -2\sqrt{\Sigma_{kk}},\partial \Omega_k)\right], C_m \cdot y_{k} \right),
\end{equation}
where $C_m=0.1$ caps the maximum measurement error standard deviation to $10\%$ of the mean data values and $C=0.2$ controls how close the noise-perturbed data can come within physical boundaries $\partial \Omega_i$ of the data (e.g., to keep relative humidities between 0 and 1 with high probability). 

\subsection{Calibrate: Ensemble Kalman Inversion}
\label{sec:calibrate}
The first stage of the CES methodology is to calibrate the parameters $\bm{\theta}$ of the model based on data $\bm{y}$.
We perform calibration with independent realizations of $\mathcal{G}(\bm{\theta}_i, \xi_i)$, viewed as noisy approximations of $\mathcal{G}_\infty(\bm{\theta}_i)$. Calibration is performed using ensemble Kalman methods, which demonstrate theoretical success, in idealized problems, and empirical success, in complex problems, to optimize parameters under such noise \cite{duncan2021ensemble}.

The utility of the calibration stage is two-fold: (1) optimize parameters to minimize the mismatch between model output and data; and (2) provide good parameter--model output pairs $(\theta_i, \mathcal{G}(\bm{\theta}_i, \xi_i))$ for training an emulator of the parameter-to-data map, with a higher density of training points near the optimal parameters.
The ensemble Kalman filter (EnKF) is a Kalman filter implementation in which the covariances are approximated using Monte Carlo sampling \cite{evensen2003ensemble}.
The EnKF has been used widely for derivative-free state estimation in numerical weather prediction (e.g., \cite{houtekamer2016review}) and model-based control (e.g., \cite{howland2020optimal}).
Ensemble Kalman methods for Bayesian inversion were introduced by \cite{chen2012ensemble} 
and \cite{emerick2013ensemble}. 
These methods provably draw samples from the posterior of linear inverse problems subject to additive Gaussian noise; however, they fail to do
so more in more general, nonlinear problems.
Recognizing this, the offline ensemble Kalman inversion (EKI) \cite{iglesias2013ensemble} algorithm was introduced for classical, optimization-based inversion.
EKI generally drives the ensemble members toward consensus near the optimal solution of the inverse problem  \cite{schillings2017analysis}.
The parameter update of ensemble member $m$ at iteration step $n$ is
\begin{equation}
\bm{\theta}_m^{(n+1)} = \bm{\theta}_m^{(n)} + C_{\bm{\theta} \mathcal{G}}^{(n)} \left( \Sigma + \Delta + C_{\mathcal{G} \mathcal{G}}^{(n)} \right)^{-1} \left( \bm{y} - \mathcal{G}(\bm{\theta}_m^{(n)}, \xi_m) \right),
\label{eq:eki}
\end{equation}
where $C_{\bm{\theta} \mathcal{G}}$ is the empirical cross-covariance between the parameters and the model outputs, and $C_{\mathcal{G} \mathcal{G}}$ is the empirical covariance of the model outputs.

EKI is guaranteed to find the optimal parameters in linear problems \cite{schillings2017analysis} and has empirical success in nonlinear problems \cite{dunbar2020calibration}.
Several other approaches exist for estimation in nonlinear problems.
\cite{li2007iterative} and \cite{sakov2012iterative} developed iterative EnKF methods which improve state estimation in strongly nonlinear problems.
However, the distribution of the ensemble does not converge to the posterior distribution in the limit of infinite members for nonlinear problems \cite{zhou2006assessing, annan2007efficient, gland2009large, garbuno2020interacting}, necessitating the emulation and sampling described in the following sections.

The number of ensemble members and EKI iterations are hyperparameters; they are set to standard values of $N_{\mathrm{ens}}=100$ and $N_{\mathrm{it}}=5$ in this study \cite{schillings2017analysis, dunbar2020calibration, cleary2021calibrate}.
Each ensemble member is run through the GCM, initialized from the same initial conditions.
A spin-up period of one year ($360$ days) is run before the statistics are computed using Eq. \eqref{eq:g_season}, to ensure the forward model realizations are subject to differing instantiations of internal variability in the chaotic GCM system.

\subsection{Emulate: Gaussian process emulation}
\label{sec:emulate}
We emulate the mapping from parameters to model output using a machine learning method that enables the rapid execution of the mapping, compared to the computationally expensive forward model.
The calibration stage (section \ref{sec:calibrate}) results in $N_t=N_{\mathrm{ens}} \cdot N_{\mathrm{it}}$ input-output pairs  $\{\bm{\theta}_i, \mathcal{G}(\bm{\theta}_i,\xi_i)\}_{i=1}^{N_{t}}$ of model parameter to model output. Harnessing the input-output pairs as training points, 
Gaussian process regression is used \cite{rasmussen2003gaussian} to create an emulator, composed of a mean function and covariance function pair, where $\mathcal{G}_{\mathrm{GP}}(\bm{\theta}) \approx \mathcal{G}_\infty(\bm{\theta})$ and $\Sigma_{\mathrm{GP}} \approx \Sigma$.
Since the input-output pairs are subject to different realizations of the chaotic system, the emulator mean approximates $\mathcal{G}_\infty(\bm{\theta})$ rather than $\mathcal{G}(\bm{\theta}, \xi)$ \cite{cleary2021calibrate, dunbar2020calibration}.

The variables of interest in the synthetic data $Y\in \mathbb{R}^{N \times N_d}$ are correlated. 
The output variables from the GCM forward model $\mathcal{G}(\bm{\theta}, \xi)$ are similarly correlated.
The correlation between the GCM statistics results in nonzero off-diagonal covariance matrix elements.
In order to maintain a diagonal covariance in the GP emulator $\Sigma_{\mathrm{GP}}$, we transform the GCM statistics into a decorrelated space using principal component analysis on $\Sigma \in \mathbb{R}^{N \times N}$ \cite{cleary2021calibrate}.
That is, we decompose the covariance matrix using the singular value decomposition
\begin{equation}
\Sigma = V D^2 V^\intercal,
\end{equation}
with a matrix of principal component vectors (or singular vectors) $V$ and the diagonal matrix $D$ containing the square roots of the singular values $\sigma_i$.
Often, in practical data assimilation problems, $N>N_d$ \cite{houtekamer2016review}. 
In this case, the covariance matrix   is rank deficient, with  $\mathrm{rank}(\Sigma)\le\min(N, N_d)$,
with singular values $\sigma_i^2=0$ for $i>\mathrm{rank}(\Sigma)$.
Methods to decorrelate the data and model output with rank deficient covariance matrices are discussed in \ref{sec:decorrelation}.

The GP is trained using input-output pairs in the decorrelated space $\{\bm{\theta}_i, \tilde{{\mathcal{G}}}(\bm{\theta}_i, \xi_i)\}_{i=1}^{N_t}$ with the decorrelated space denoted by tildes, $\tilde{(\cdot)}$.
The resulting input-output mapping is approximated as
\begin{equation}
\tilde{{\mathcal{G}}}_\infty(\bm{\theta}) \approx N(\tilde{{\mathcal{G}}}_{\mathrm{GP}}(\bm{\theta}), \tilde{{\Sigma}}_{\mathrm{GP}}(\bm{\theta})).
\end{equation}
The GP kernels used are a white-noise kernel added to an Automatic Relevance Determination (ARD) radial basis function kernel; further details provided in \cite{dunbar2020calibration, cleary2021calibrate}.
The GP hyperparameters are trained using the input-output pairs.
The training process results in a GP regression function $\tilde{{\mathcal{G}}}_{\mathrm{GP}}(\bm{\theta})$ which takes $\bm{\theta}$ as input and emulates $\tilde{{\mathcal{G}}}_\infty(\bm{\theta})$ in a computationally efficient fashion.

\subsection{Sample: Posterior sampling using MCMC}
\label{sec:sample}

The Bayesian posterior distribution is approximated through MCMC sampling with the trained GP emulator $\tilde{{\mathcal{G}}}_{\mathrm{GP}}(\bm{\theta})$.
The data $\bm{y}$ is normalized and transformed into the decorrelated space $\tilde{{\bm{y}}}$, as in section \ref{sec:emulate}.
The Bayesian posterior distribution is approximated as \cite{stuart2010inverse}
\begin{align}
\mathbb{P}(\bm{\theta} | \tilde{{\bm{y}}}) &\propto 
\mathbb{P}(\tilde{{\bm{y}}} | \bm{\theta}) \mathbb{P}(\bm{\theta}) \\
&\propto \exp \left( 
-\frac{1}{2} \| \tilde{{\bm{y}}}- \tilde{{\mathcal{G}}}_{\mathrm{GP}}(\bm{\theta}) \|_{\tilde{{\Sigma}}_{\mathrm{GP}}(\bm{\theta}) + \tilde{{\Delta}} }^2
-\frac{1}{2} \log \det \left( \tilde{{\Sigma}}_{\mathrm{GP}}(\bm{\theta}) + \tilde{{\Delta}} \right)
\right) 
\mathbb{P}(\bm{\theta}).
\end{align}
With a normal prior distribution governed by mean $\overline{\bm{\theta}}$ and covariance $\Gamma_{\bm{\theta}}$ the resulting MCMC objective function is
\begin{equation}
\Phi_{\mathrm{MCMC}}(\bm{\theta},\tilde{{\bm{y}}}) = \exp \left( 
-\frac{1}{2} \| \tilde{{\bm{y}}} - \tilde{{\mathcal{G}}}_{\mathrm{GP}}(\bm{\theta}) \|_{\tilde{{\Sigma}}_{\mathrm{GP}}(\bm{\theta}) + \tilde{{\Delta}}}^2 
-\frac{1}{2} \log \det \left( \tilde{{\Sigma}}_{\mathrm{GP}}(\bm{\theta}) + \tilde{{\Delta}} \right) 
-\frac{1}{2} \| \bm{\theta} - \overline{\bm{\theta}} \|_{\Gamma_{\bm{\theta}}}^2
\right).
\end{equation}
We use a random walk Metropolis MCMC algorithm. 
The number of MCMC samples is set to $N_{\mathrm{MCMC}}=200,000$ with a burn-in of $10,000$.


\section{Seasonal GCM uncertainty quantification}
\label{sec:gcm}

We perform numerical experiments with an idealized GCM with a seasonal cycle.
The GCM simulation setup is presented in section \ref{sec:gcm_setup}.
Various climate statistics that we extract from the GCM and use in the numerical experiments are discussed in section \ref{sec:stats} and shown in section \ref{sec:truth}.

\subsection{Seasonal simulation setup}
\label{sec:gcm_setup}

The GCM used in this study is based on the Geophysical Fluid Dynamics Laboratory's Flexible Modeling System \cite{frierson2006gray}.
The GCM simulates an idealized aquaplanet with a homogeneous mixed-layer slab ocean bottom boundary condition with a depth of $1~\mathrm{m}$.
The GCM has been used previously for simulations of the hydrological cycle over a range of climates \cite{o2008hydrological} and to characterize seasonal variability in the tropics \cite{bordoni2008monsoons,Merlis13a,Merlis13c,kaspi2011winter,Bischoff14a,Bischoff16a,wei2018energetic}. 
The GCM is axisymmetric and statistically cyclostationary.
The spectral transform method is used in the horizontal directions, and finite differencing in sigma coordinates is used in the vertical direction.
The horizontal resolution used is T21 with $N_\phi=32$ discrete latitude points on the transform grid, and $20$ sigma levels ($\sigma_p=p/p_s$, where $p$ is the pressure and $p_s$ is the local surface pressure).

A two-stream gray radiation scheme is used.
The top-of-atmosphere (TOA) insolation is prescribed and varies according to a seasonal cycle \cite{bordoni2008monsoons, wei2018energetic}.
The diurnal cycle insolation variations are neglected, with a daily average insolation applied.
The longwave and shortwave optical thicknesses depend on the latitude and pressure.
The radiative effects of variations of atmospheric water vapor or clouds are neglected, and therefore, water vapor and cloud feedbacks are not included in the GCM. 

The convection is parameterized using a simplified quasi-equilibrium Betts-Miller (SBM) scheme \cite{betts1986new, betts1986new2,frierson2007dynamics}.
Vertical profiles of temperature and humidity are used to calculate precipitation and associated temperature and humidity changes through a relaxation to moist adiabatic reference profiles \cite{frierson2007dynamics}.
The relaxation is included as a forcing to the temperature and humidity balances
\begin{align}
\frac{\partial T}{\partial t} + \cdots &= S_T - f_T \frac{T-T_{\mathrm{ref}}}{\tau} \\
\frac{\partial q}{\partial t} + \cdots &= S_q - f_q f_T \frac{q-q_{\mathrm{ref}}}{\tau},
\end{align}
where the dots represent the dynamical terms in the equations, and $S_T$ and $S_q$ represent the temperature and specific humidity forcings aside from the convection scheme.
The term $f_T$ governs the spatiotemporal activation of the convection scheme, depends on the thermodynamic state, and is a function of $z$.
The term $f_q$ modifies the specific humidity relaxation \cite{o2008hydrological}.
The reference temperature $T_{\mathrm{ref}}$ is a moist adiabat, chosen so that the convection scheme conserves enthalpy integrated over vertical columns \cite{frierson2007dynamics,o2008hydrological}.
The reference specific humidity $q_{\mathrm{ref}}$ is that which corresponds to a prescribed relative humidity with respect to the moist adiabat $T_{\mathrm{ref}}$.
For the UQ experiments, our focus are two parameters: the prescribed reference relative humidity parameter ($\theta_{\mathrm{RH}}$) and the relaxation timescale parameter ($\theta_\tau$).

The GCM simulation starts from vernal equinox, and the year length is $T_{yr} = 360~\mathrm{d}$.
All climate statistics used for UQ are zonally averaged. We use the Betts-Miller convection scheme with standard reference values of the parameters of $\bm{\theta}^\dagger=(\theta_{\mathrm{RH}}^\dagger, \theta_{\tau}^\dagger) = (0.7, 7200~\mathrm{s})$ \cite{frierson2007dynamics, o2008hydrological}.
UQ of $\bm{\theta}$ relies on a prior knowledge about the convective parameters.
We use wide prior distributions of the parameters, which enforce physical constraints, such as $0< \theta_\mathrm{RH} \le 1$ and $\theta_\tau > 0$, but are otherwise uninformative \cite{dunbar2020calibration}.
The selected priors are $\theta_{\mathrm{RH}} \sim \mathrm{Logit}(N(0,1))$ and $\theta_{\tau} \sim \mathrm{Log}(N(12 \mathrm{h},(12 \mathrm{h})^2))$, that is, normal distributions of logit- and log-transformed parameters.
The parameter priors are independent, although joint prior distributions could be used in future work.

\subsection{Climate statistics used for UQ}
\label{sec:stats}

We calibrate and perform UQ on the parameters of the convection scheme using more and less informative time-averaged data from GCM simulations. 
The informative statistics, such as the mid-tropospheric relative humidity, are chosen because they are strongly affected by the choice of the convective parameters (e.g., \cite{o2008hydrological,schneider2008moist,OGorman11b}).
For comparison, we also choose less informative statistics that are affected less by the choice of convective parameters, such as surface wind speeds.
Numerical UQ experiments using the differing degrees of information in the climate statistics will illustrate the impact of incorporating higher frequencies in the climate statistics used for parameter estimation.

\begin{table}
\def~{\hphantom{0}}
  \begin{tabular}{c|p{65mm} p{45mm}}
  \toprule
     & Informative & Less informative \\
    \hline
    Variables & 1. Relative humidity ($\sigma_p=0.5$) \newline 2. Precipitation rate (mm/day) \newline 3. Probability of $90^{th}$ percentile precipitation
    & 1. Precipitation rate (mm/day) \newline 2. Surface wind speed (m/s) \\
    \hline
    Latitudes, $N_\phi$ & $32$ & $32$  \\
    $N_y$ & $96$ & $64$  \\
    $N, \ T=360~\mathrm{d}$ (annual) & $96$ & $64$  \\
    $N, \ T=90~\mathrm{d}$ (seasonal) & $384$ & $256$  \\
    \toprule
  \end{tabular}
  \caption{Summary of the climate data used.
  Calibration and UQ is performed using data that are more or less informative about the convective scheme. The total size of the data used for UQ is $N$.
  }
  \label{tab:statistics}
\end{table}

For the informative climate statistics, we choose three variables:
the mid-tropospheric ($\sigma_p=0.5$) relative humidity, the total precipitation rate, and, as a measure of precipitation intensity, the probability that a daily precipitation total exceeds the latitude-dependent 90th precipitation percentile threshold from a long control simulation with the true parameters \cite{dunbar2020calibration}.
Since intense precipitation events are influenced by the convection scheme \cite{o2009physical, o2009scaling}, exceedances of precipitation over a high threshold are anticipated to be informative about the convection scheme parameters.
The three statistics, are evaluated at each of the $N_\phi=32$ discrete latitudes, giving $96$ total quantities of interest.
For less informative statistics, we use the zonally averaged precipitation rate and surface wind speed.
As with the informative statistics, $N_\phi=32$ discrete latitudes are considered, giving $64$ total quantities of interest (see Table \ref{tab:statistics} for a summary).

We nondimensionalize the GCM statistics (Eq. \eqref{eq:normalization}) with the median $\bm{y}_c$, taken over latitude and time, of each specified data type.
There is one characteristic value for each data type, e.g., one characteristic value for precipitation and a separate characteristic value for relative humidity.

\subsection{Synthetic data used for UQ}
\label{sec:truth}

\begin{figure}
    \centering
    \includegraphics[width=\linewidth]{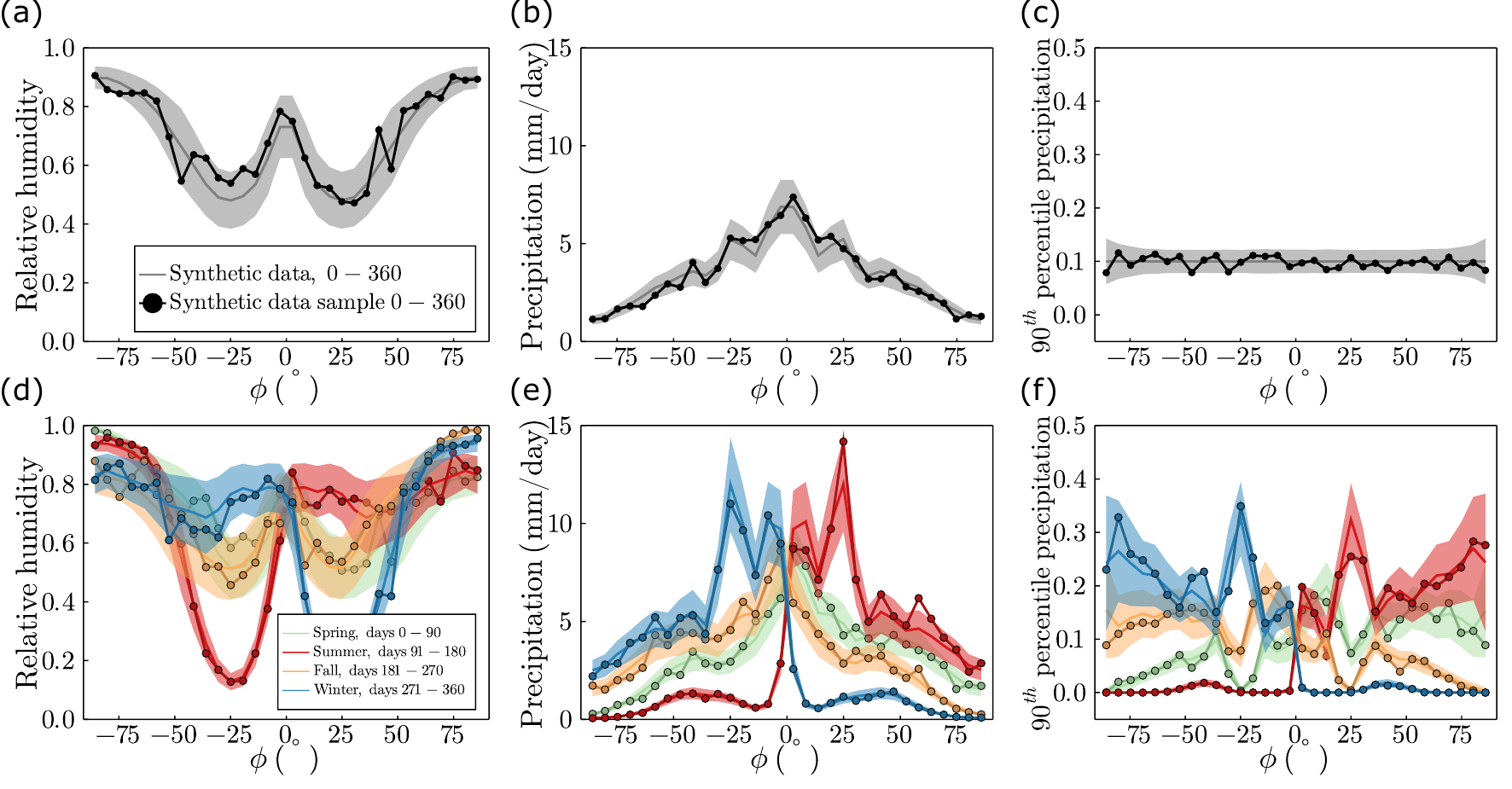}
    \caption{Informative synthetic data with
  (a--c) $T=360~\mathrm{d}$ and (d--f) $T=90~\mathrm{d}$.
  The light solid lines correspond to $\bm{y}_{\infty}$.
  The shaded regions correspond to a $95\%$ confidence interval around $\bm{y}_{\infty}$ with covariance $\Sigma + \Delta$.
  The dark solid lines with circle markers correspond to a randomly selected sample of the synthetic data $\bm{y}$ that has been subjected to measurement noise $N(0,\Delta)$.
  Day $0$ corresponds to vernal equinox and the Northern hemisphere seasons are provided.
  }
    \label{fig:truth_inflate_1}
\end{figure}

\begin{figure}
    \centering
    \includegraphics[width=0.66\linewidth]{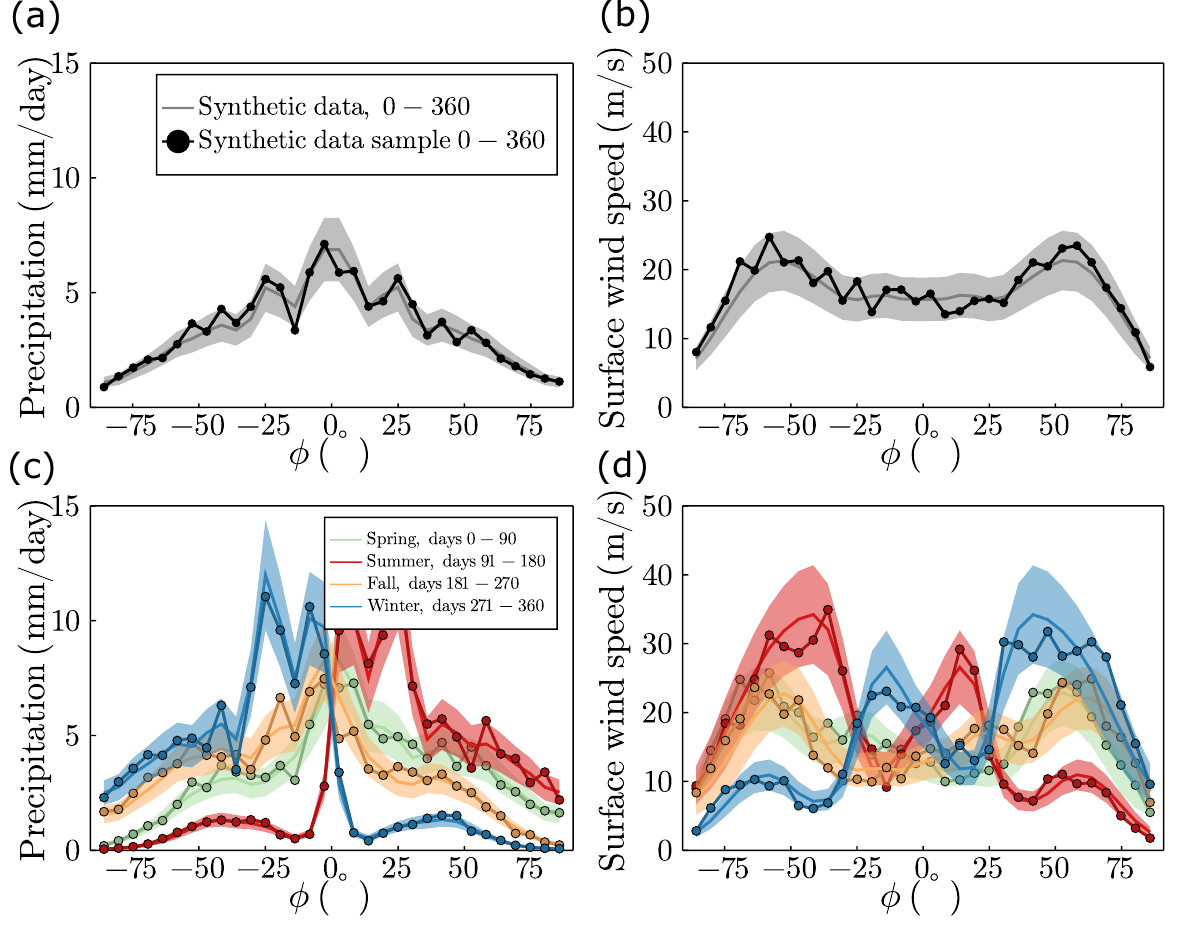}
    \caption{Less informative synthetic data with (a, b) $T=360~\mathrm{d}$ and (c, d) $T=90~\mathrm{d}$.
  The light solid lines correspond to $\bm{y}_{\infty}$.
  The shaded regions correspond to a $95\%$ confidence interval around $\bm{y}_{\infty}$ with covariance $\Sigma + \Delta$.
  The dark solid lines with circle markers correspond to a randomly selected sample of the synthetic data $\bm{y}$ that has been subjected to measurement noise $N(0,\Delta)$.
  Day $0$ corresponds to vernal equinox and the Northern hemisphere seasons are provided.}
    \label{fig:wind_truth_inflate_1}
\end{figure}

As synthetic data, we used GCM output with different initial conditions, and perturbed with measurement error with covariance $\Delta$.
The synthetic states are constructed by running the seasonal GCM for $150$ years at the true parameters $\bm{\theta}^\dagger$.
For each case of $T=90~\mathrm{d}$ and $T=360~\mathrm{d}$, the GCM states are aggregated according to the method introduced in section \ref{sec:seasonal_inverse}.
The informative synthetic statistics for $T=360~\mathrm{d}$ days are shown in Figure \ref{fig:truth_inflate_1}a--c, which compares the $N_d=150$ year ensemble average of the statistics with a $1$ year sample subject to internal variability and perturbed with measurement noise $N(0,\Delta)$.
The synthetic data for $T=90~\mathrm{d}$ are shown in Figure \ref{fig:truth_inflate_1}d--f.
For the data averaged over $90~\mathrm{d}$, a distinct seasonal cycle emerges with, for example, relatively high precipitation and relative humidity in the northern hemisphere summer (days 90--180).
The less informative statistics of precipitation rate and surface wind speed are shown for $T=360~\mathrm{d}$ and $T=90~\mathrm{d}$ in Figure \ref{fig:wind_truth_inflate_1}.

\section{GCM calibration and UQ results}
\label{sec:results}


\subsection{Informative statistics}
\label{sec:informative}


The convective parameters $\theta_{\mathrm{RH}}$ and $\theta_{\tau}$ are calibrated using EKI with the informative statistics with $T=90~\mathrm{d}$ and $T=360~\mathrm{d}$ (see Table \ref{tab:statistics}).
The EKI calibration is performed with a synthetic data sample $\bm{y}$ (Eq. \ref{eq:inverse_initial_differ}).
Since the synthetic data is subject to internal variability, the corresponding calibration is influenced by the synthetic sample values.
The synthetic data sample is randomly selected from the $150$ years of historical data constructed with the true convective parameters $\bm{\theta}^\dagger$ (see section \ref{sec:stats}).
Calibration is performed with $10$ independent synthetic samples, and the results in this section are presented as means and standard deviations over the $10$ independent realizations.
The cases with $T=90~\mathrm{d}$ and $T=360~\mathrm{d}$ are run with the same $10$ synthetic samples.

\begin{figure}
  \centering
  \begin{tabular}{@{}p{0.5\linewidth}@{\quad}p{0.5\linewidth}@{}}
  \subfigimgthree[width=\linewidth,valign=t]{(a)}{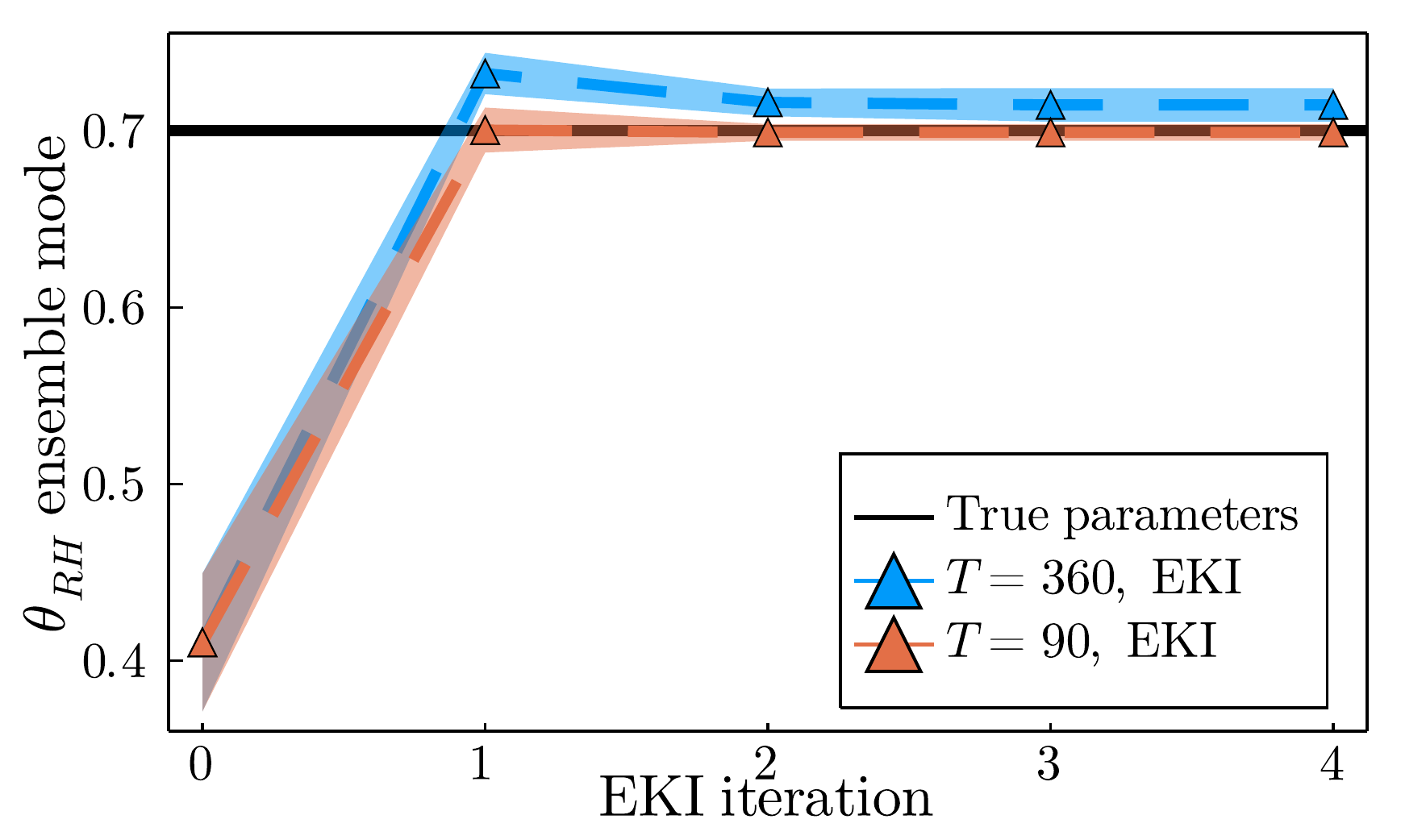} &
  \subfigimgthree[width=\linewidth,valign=t]{(b)}{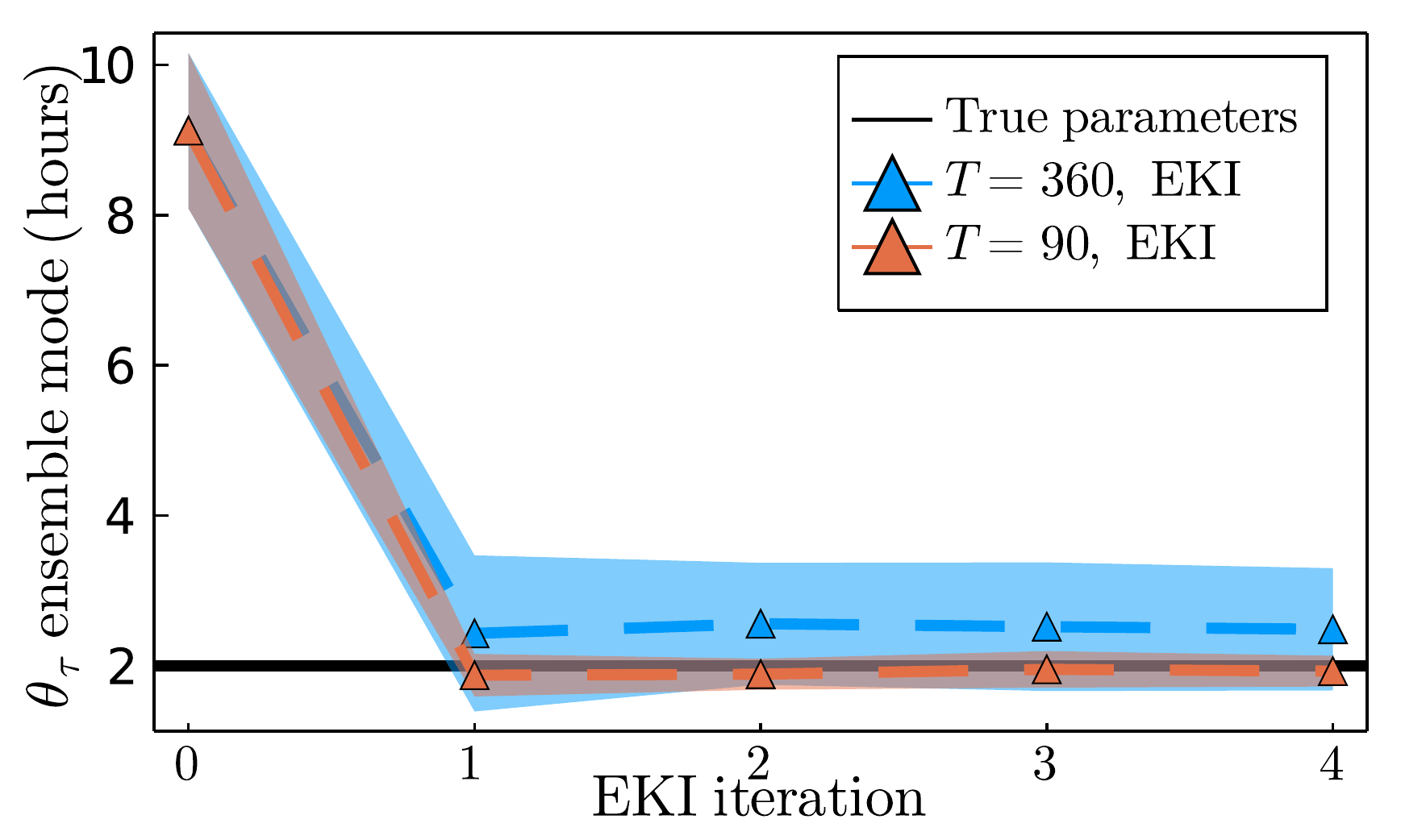}
  \end{tabular}
  \caption{Ensemble Kalman inversion convective parameter estimates using informative statistics. 
  Mode of (a) the estimated relative humidity convective parameter $\theta_{\mathrm{RH}}$ compared to truth value $\theta_{\mathrm{RH}}^\dagger$ and of (b) the estimated relaxation time scale convective parameter $\theta_{\tau}$ compared to truth value $\theta_{\tau}^\dagger$.
  The mean and standard deviation over independent instantiations with differing synthetic samples are shown by the lines and shaded regions, respectively.}
    \label{fig:ens_of_ens_modes}
\end{figure}

The modes of the EKI ensemble distributions are shown in Figure \ref{fig:ens_of_ens_modes}.
The modes demonstrate a reduction in estimation error of the true convective parameters when the data are aggregated seasonally with $T=90~\mathrm{d}$.
The calibrated convective parameters using annually averaged GCM statistics have higher mean bias and variability between the instantiations with different synthetic data samples.
For both integration timescales, the error associated with the $\theta_\tau$ estimation was larger than the estimation error for $\theta_{\mathrm{RH}}$.
The percent error of the average mode at the last EKI iteration for $\theta_{\mathrm{RH}}$ is $0.3\%$ and $0.8\%$ for $T=90~\mathrm{d}$ and $T=360~\mathrm{d}$, respectively.
For $\theta_\tau$, the percent errors are $1.2\%$ and $11.5\%$ for $T=90~\mathrm{d}$ and $T=360~\mathrm{d}$, respectively.
This indicates that the integration of the GCM states over an annual cycle filters information from the resulting GCM statistics that is informative, especially about the relaxation timescale.

The normalized mean square error (MSE) for $\theta_{\mathrm{RH}}$ at EKI iteration $(n)$ is
\begin{equation}
\varepsilon^{(n)}_{\mathrm{RH}} = \frac{1}{{\theta^\dagger_{\mathrm{RH}}}} \sqrt{ \frac{1}{N_{\mathrm{ens}}} \sum_{i=1}^{N_{\mathrm{ens}}} \left(\theta_{\mathrm{RH},i} - \theta_{\mathrm{RH}}^\dagger \right)^2}.
\end{equation} 
The error is computed similarly for $\theta_\tau$.
For both the relative humidity and relaxation timescale parameters, $T=90~\mathrm{d}$ reduces the parameter estimation MSE relative to $T=360~\mathrm{d}$ by about a factor 2--3: The MSE for $\theta_{\mathrm{RH}}$ is $0.006$ and $0.015$ for $T=90~\mathrm{d}$ and $T=360~\mathrm{d}$, respectively; 
for $\theta_\tau$, the MSE is $0.06$ and $0.23$ for $T=90~\mathrm{d}$ and $T=360~\mathrm{d}$, respectively.
The seasonally aggregated data also have a smaller spread among the realizations, as visualized by the uncertainty bands around the mean values.
The reductions in the MSE standard deviations were $43\%$ and $66\%$ for $\theta_{\mathrm{RH}}$ and $\theta_\tau$, respectively.

\begin{figure}
    \centering
    \includegraphics[width=\linewidth]{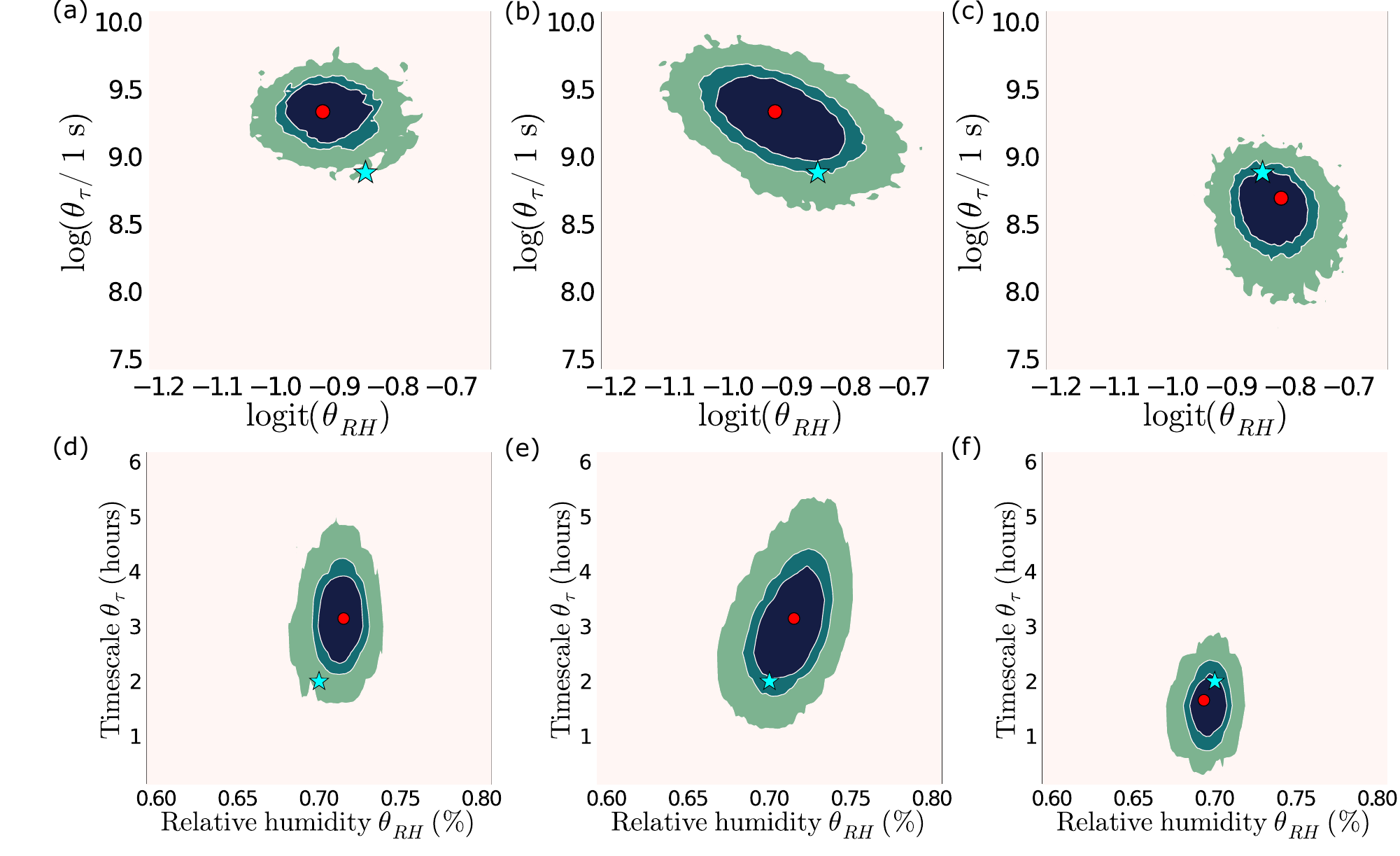}
    \caption{
  Convective parameter posterior distributions computed using MCMC using a GP emulator that was trained using data aggregated with an integration timescale of $T=360~\mathrm{d}$ and $T=90~\mathrm{d}$. The contours correspond to $50\%$, $75\%$, and $99\%$ of the posterior distribution, the star is the truth, and the circle is the average of the ensemble members after the last EKI iteration. Posteriors are shown for (a) $T=360~\mathrm{d}$, decorrelated with full SVD, and (b) a 95\%-energy truncation of the SVD. 
  (c) Posterior for $T=90~\mathrm{d}$, decorrelated with 95\%-energy truncated SVD.
  Panels (d--f) same as (a--c) with the posterior distributions shown in the physical parameter space.
  }
    \label{fig:mcmc_inflation_1_orig_90_vs_360}
\end{figure}

The input-output pairs generated during EKI parameter calibration are used for training the GP emulator.
These pairs of parameters and model evaluations are taken from four 100-member EKI iterations as well the initial 100-member ensemble drawn from the prior, giving a total of 500 input-output pairs for GP training.
The results of this study were not significantly different with more training points.
The GCM outputs are mapped into a normalized and decorrelated space according to the SVD of the internal variability covariance matrix $\Sigma$, truncated to contain $95\%$ of the energy in the singular values.
For $T=90~\mathrm{d}$, the truncation contains $k=47$ singular values, while $T=360~\mathrm{d}$ retains $k=23$ (see \ref{sec:trunc_vs_reg}). 
A scalar-valued GP is trained for each of the $k$ outputs.
For comparison, we also ran the emulation and sampling steps for $T=360~\mathrm{d}$ without SVD truncation.

The posterior distributions are approximated using MCMC sampling with the trained GP emulators.
Posterior distributions for the two convective parameters for a randomly selected synthetic data sample are shown in Figure \ref{fig:mcmc_inflation_1_orig_90_vs_360}.
In Figure \ref{fig:mcmc_inflation_1_orig_90_vs_360}a, the posterior distribution was computed with normalized input data, aggregated annually with $T=360~\mathrm{d}$, but the SVD was not truncated before emulation and sampling.
The posterior resulting from the same initial ensemble and synthetic data sample but with SVD truncation is shown in Figure \ref{fig:mcmc_inflation_1_orig_90_vs_360}b.
The truncation of the SVD smooths and inflates the posterior distribution.
While the posteriors are qualitatively similar, they differ quantitatively.
The true parameters are outside the region containing $99\%$ of the posterior mass for the untruncated case but are within the region containing $75\%$ of the posterior mass for the truncated case.
Further investigation of the influence of SVD regularization on the posteriors is in \ref{sec:trunc_vs_reg}.
Hereafter, all posterior results will focus on the posterior distributions estimated using emulation and sampling in the truncated SVD space (see \ref{sec:decorrelation}).
The convective parameter posterior distribution using seasonally aggregated data with $T=90~\mathrm{d}$ with the SVD truncated at $95\%$ of the total energy is shown in Figure \ref{fig:mcmc_inflation_1_orig_90_vs_360}c.
As in the calibration stage, the posterior mode for $T=90~\mathrm{d}$ is closer to the true parameters than for $T=360~\mathrm{d}$.
Compared to the posterior with $T=360~\mathrm{d}$, the $T=90~\mathrm{d}$ posterior is also more compact.

\begin{table}
  \begin{center}
\def~{\hphantom{0}}
  \begin{tabular}{c|cc}
  \toprule
    Parameter space & $T=90~\mathrm{d}$ & $T=360~\mathrm{d}$ \\
    \hline
    Informative data & $0.5\% \pm 0.1\%$ & $1.3\% \pm 0.3\%$ \\
    Less informative data & $5.5\%  \pm 0.9\%$ & $26.0\% \pm 7.0\%$ \\
    \toprule
  \end{tabular}
  \caption{
  The ratio ($\%$) of the area occupied by the posterior to the area occupied by the prior.
  The area of the convex hull containing $75\%$ of the posterior mass for each value of $T$ and data type is normalized by the area containing $75\%$ of the prior mass.
  The means and standard deviations over independent synthetic data realizations are provided.
  Ten and four realizations are used for the informative and less informative data, respectively.
  }
  \label{tab:T_360_informative}
  \end{center}
\end{table}

We quantify the size of the posterior distributions for the two integration timescales by computing the two-dimensional area of the convex hull containing $75\%$ of the posterior mass (middle contour level in Figure \ref{fig:mcmc_inflation_1_orig_90_vs_360}).
The area containing $75\%$ of the posterior mass for each value of $T$ is normalized by the area containing $75\%$ of the prior mass for reference.
The percentages of the prior area taken up by the posteriors for $T=90~\mathrm{d}$ and $T=360~\mathrm{d}$, averaged over the $10$ CES instantiations, are shown in Table \ref{tab:T_360_informative}.
The areas are computed in the $\mathrm{logit}(\theta_{\mathrm{RH}})-\mathrm{log}(\theta_{\tau}/\mathrm{1~s})$ transformed space, in which the sampling is performed.
The area of the posterior distribution resulting from the seasonal integration is approximately a factor 2 smaller than the posterior from the annually integrated data. 
However, compared to the wide prior, both posteriors are tight, with the posterior areas occupying around $1\%$ of the prior area.


\subsection{Less informative statistics}
\label{sec:uninformative}

In many UQ applications, intuition about which quantities of interest will lead to improved parameter estimation is not available before experimentation.
While the idealized GCM and convective parameterization used in this study have a rich set of investigations to provide prior knowledge about the relationship between $\bm{\theta}$ and $\mathcal{G}(\bm{\theta})$, we also consider UQ of the parameters using less informative GCM statistics.
To do so, we use the precipitation rate and surface wind speed, zonally averaged and integrated with $T=360~\mathrm{d}$ and $T=90~\mathrm{d}$ (Figure \ref{fig:wind_truth_inflate_1}).
Four instantiations of CES with differing synthetic data samples are run for each integration length of $T=90~\mathrm{d}$ and $T=360~\mathrm{d}$.

\begin{figure}
  \centering
  \begin{tabular}{@{}p{0.5\linewidth}@{\quad}p{0.5\linewidth}@{}}
  \subfigimgtwo[width=\linewidth,valign=t]{(a)}{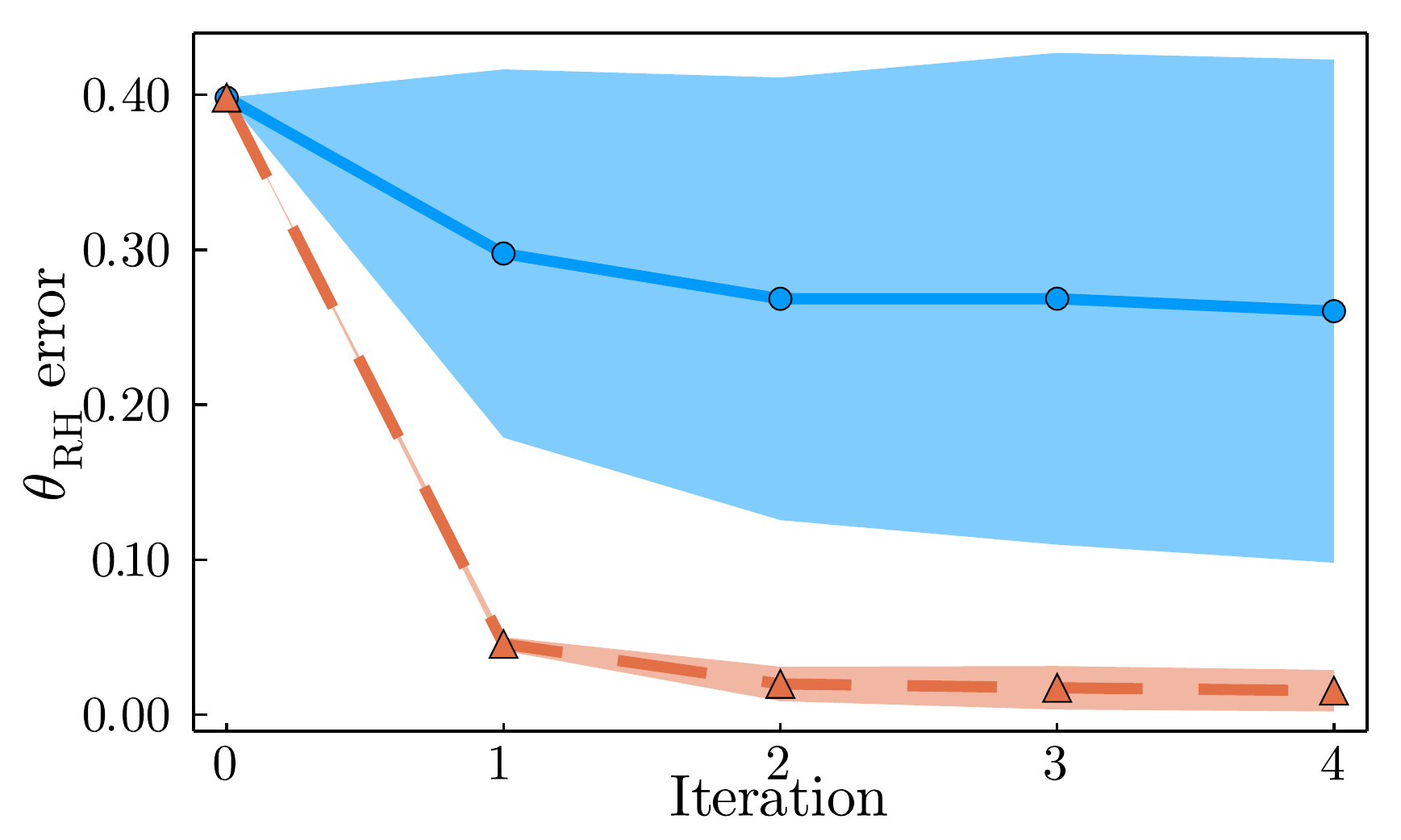} &
  \subfigimgtwo[width=\linewidth,valign=t]{(b)}{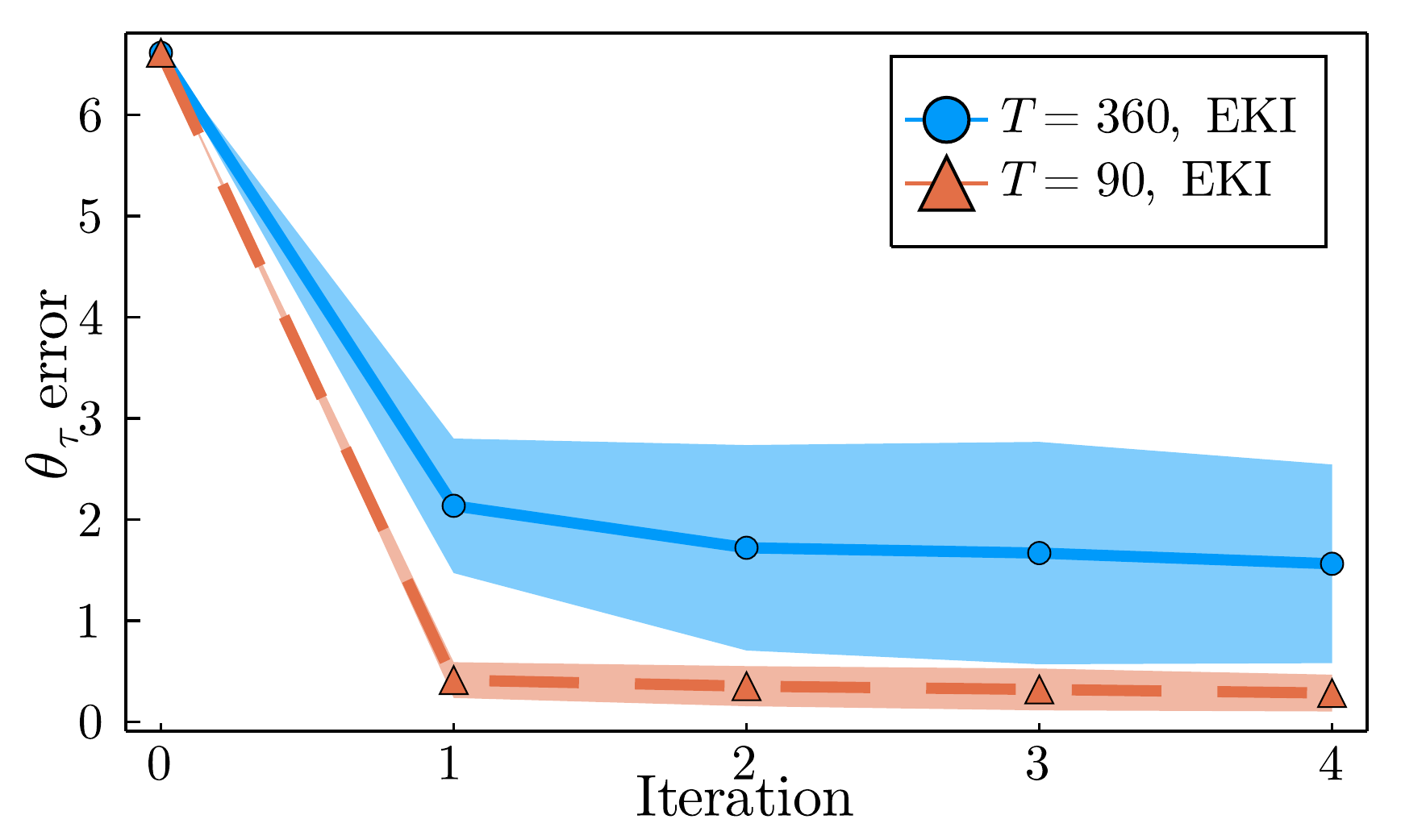}
  \end{tabular}
  \caption{
  Parameter calibration performed with EKI using less informative statistics.
  Mean square error of (a) the estimated relative humidity parameter $\theta_{\mathrm{RH}}$ compared to truth value $\theta_{\mathrm{RH}}^\dagger$ and of (b) the estimated relaxation time scale  $\theta_{\tau}$ compared to truth value $\theta_{\tau}^\dagger$.
  Solid lines are $T=360~\mathrm{d}$ and dashed lines are $T=90~\mathrm{d}$.}
    \label{fig:wind_ens_of_ens}
\end{figure}

\begin{figure}
  \centering
  \begin{tabular}{@{}p{0.5\linewidth}@{\quad}p{0.5\linewidth}@{}}
  \subfigimgthree[width=\linewidth,valign=t]{(a)}{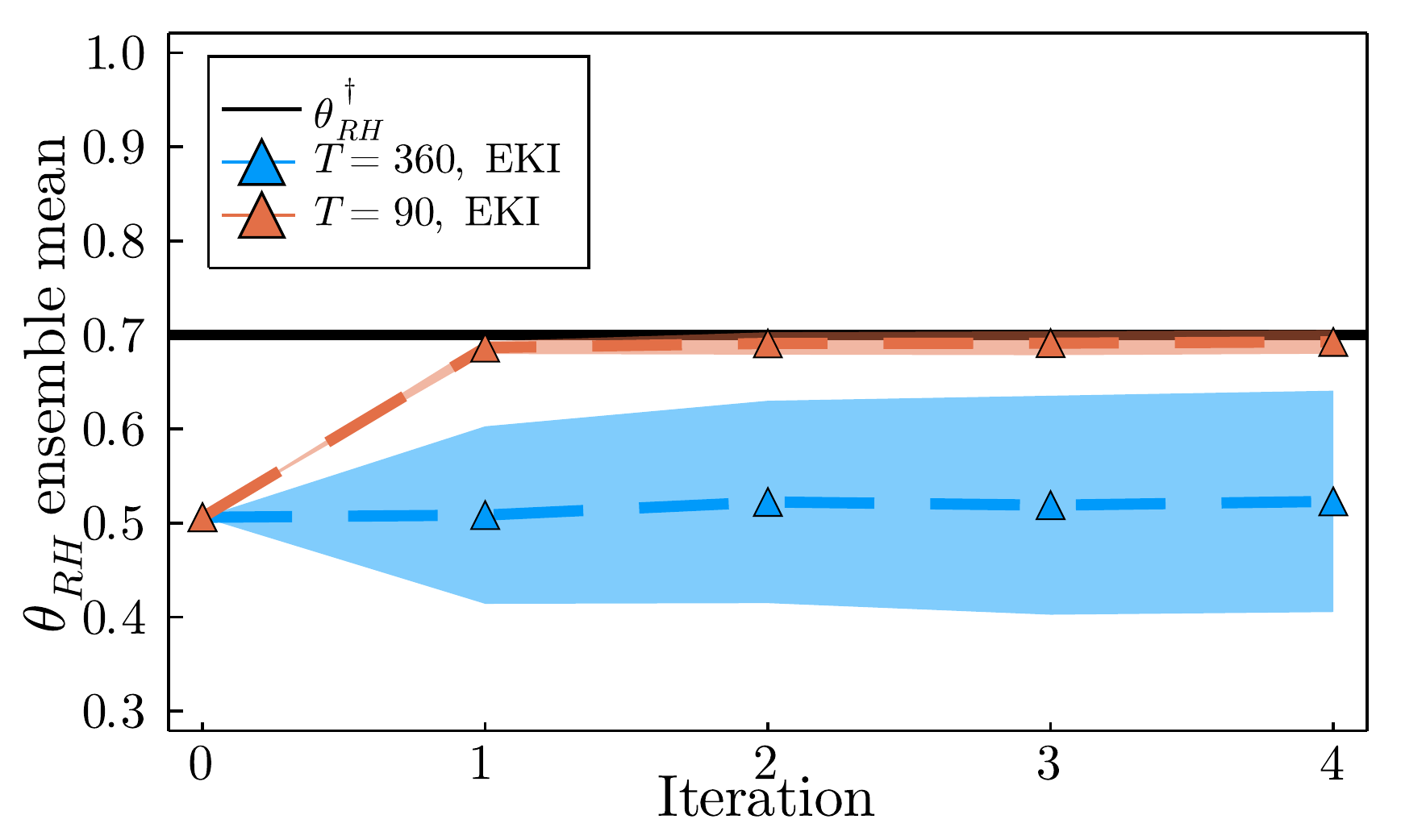} &
  \subfigimgthree[width=\linewidth,valign=t]{(b)}{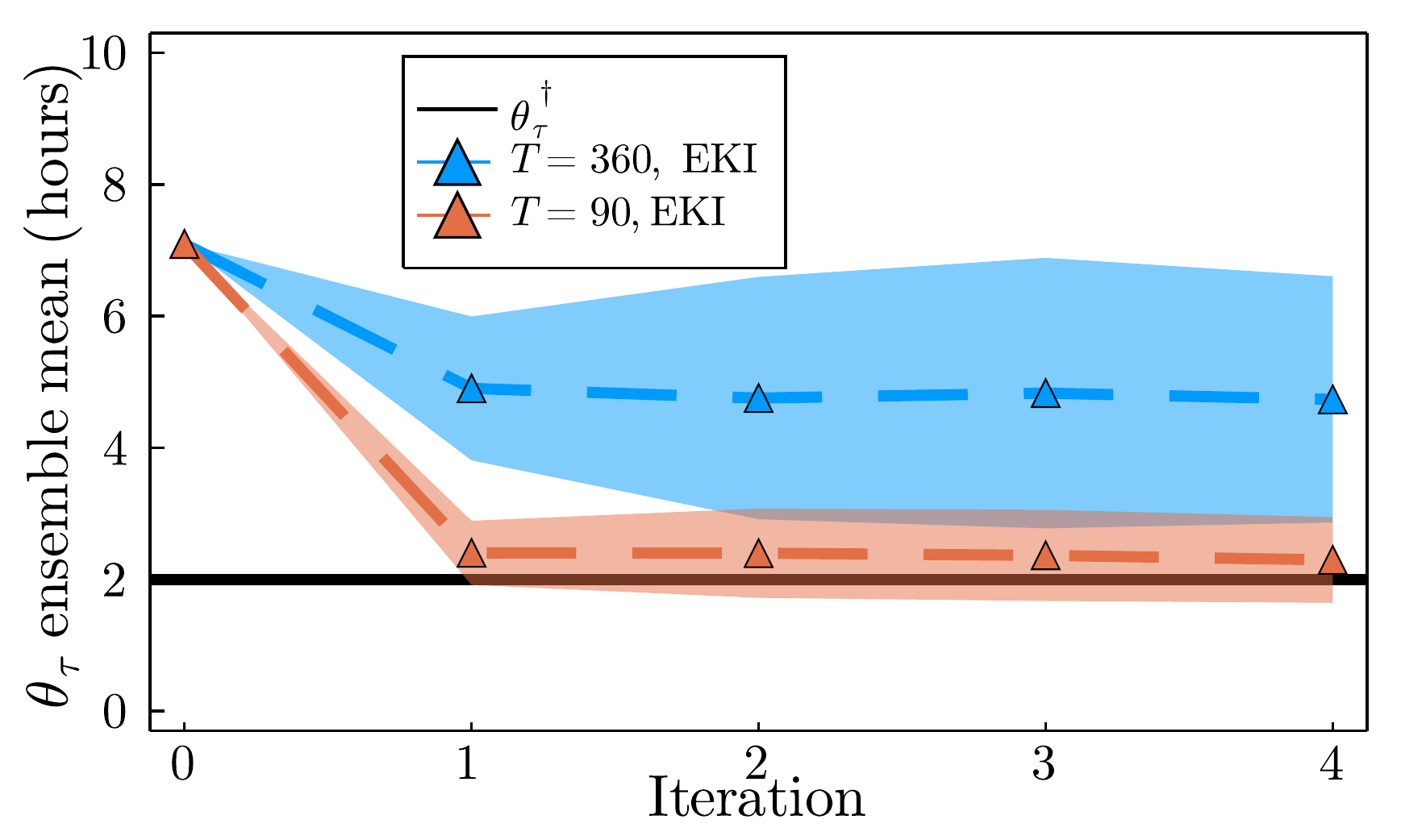} \\
  \subfigimgthree[width=\linewidth,valign=t]{(c)}{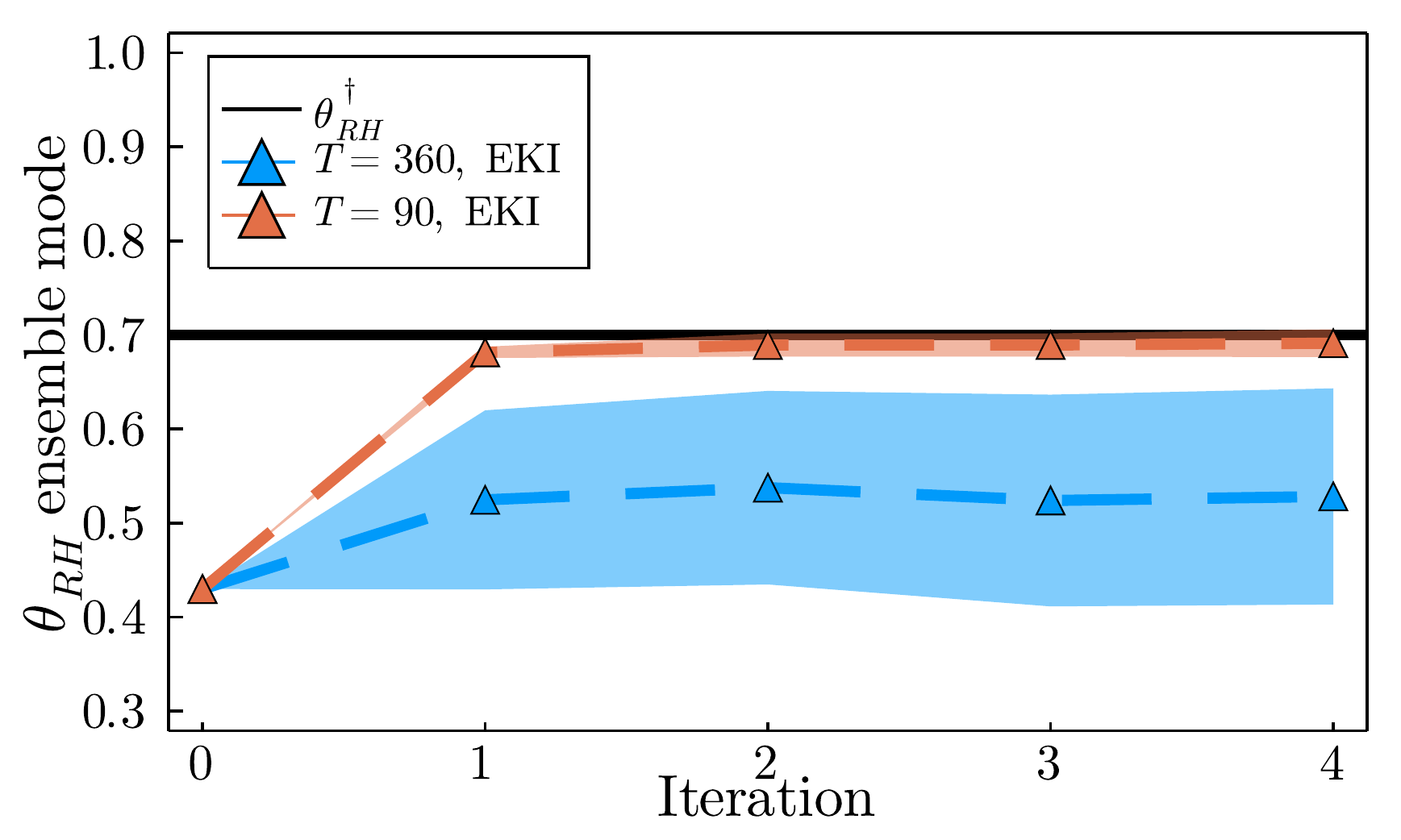} &
  \subfigimgthree[width=\linewidth,valign=t]{(d)}{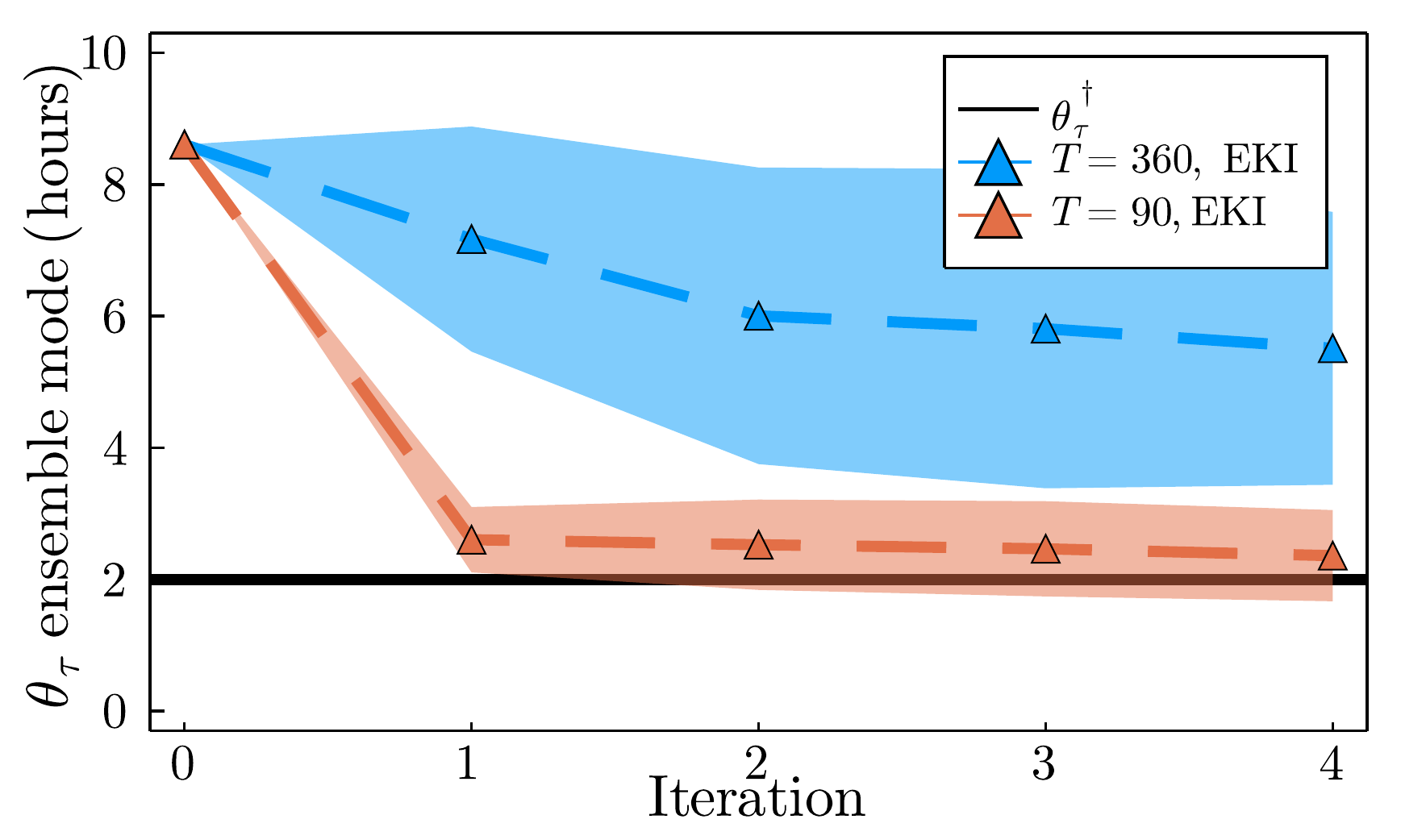} \\
  \end{tabular}
  \caption{Parameter calibration performed with EKI using less informative statistics.
  Mean of (a) the estimated relative humidity parameter $\theta_{\mathrm{RH}}$ compared to truth value $\theta_{\mathrm{RH}}^\dagger$ and of (b) the estimated relaxation time scale  parameter $\theta_{\tau}$ compared to truth value $\theta_{\tau}^\dagger$.
  (c, d) Same as (a, b) for the modes of the ensemble members.
  }
    \label{fig:wind_eki}
\end{figure}

The MSE of the EKI parameter calibration with the less informative GCM statistics are shown in Figure \ref{fig:wind_ens_of_ens} for $T=90~\mathrm{d}$ and $T=360~\mathrm{d}$.
With precipitation rate and surface wind speed as the basis for UQ, the annually averaged data result in significantly higher parameter estimation error than seasonally aggregated data.
The $T=360~\mathrm{d}$ MSEs are $0.26$ and $1.56$ for $\theta_{\mathrm{RH}}$ and $\theta_\tau$, respectively.
The $T=90~\mathrm{d}$ MSEs are $0.02$ and $0.28$ for $\theta_{\mathrm{RH}}$ and $\theta_\tau$, respectively.
The ensemble means and modes of the two parameters are shown in Figure \ref{fig:wind_eki}.
In \ref{sec:filtering}, the calibration is performed by considering the seasonal data sequentially, rather than collectively, to restrict the data size for both $T=90~\mathrm{d}$ and $T=360~\mathrm{d}$ to $N=64$.
The results demonstrate that the reductions in MSE are the result of the incorporation of seasonal information in the data, rather than the dimensionality of the data space.


\begin{figure}
  \centering
  \begin{tabular}{@{}p{0.33\linewidth}@{\quad}p{0.33\linewidth}@{}}
  \subfigimgtwo[width=\linewidth,valign=t]{(a)}{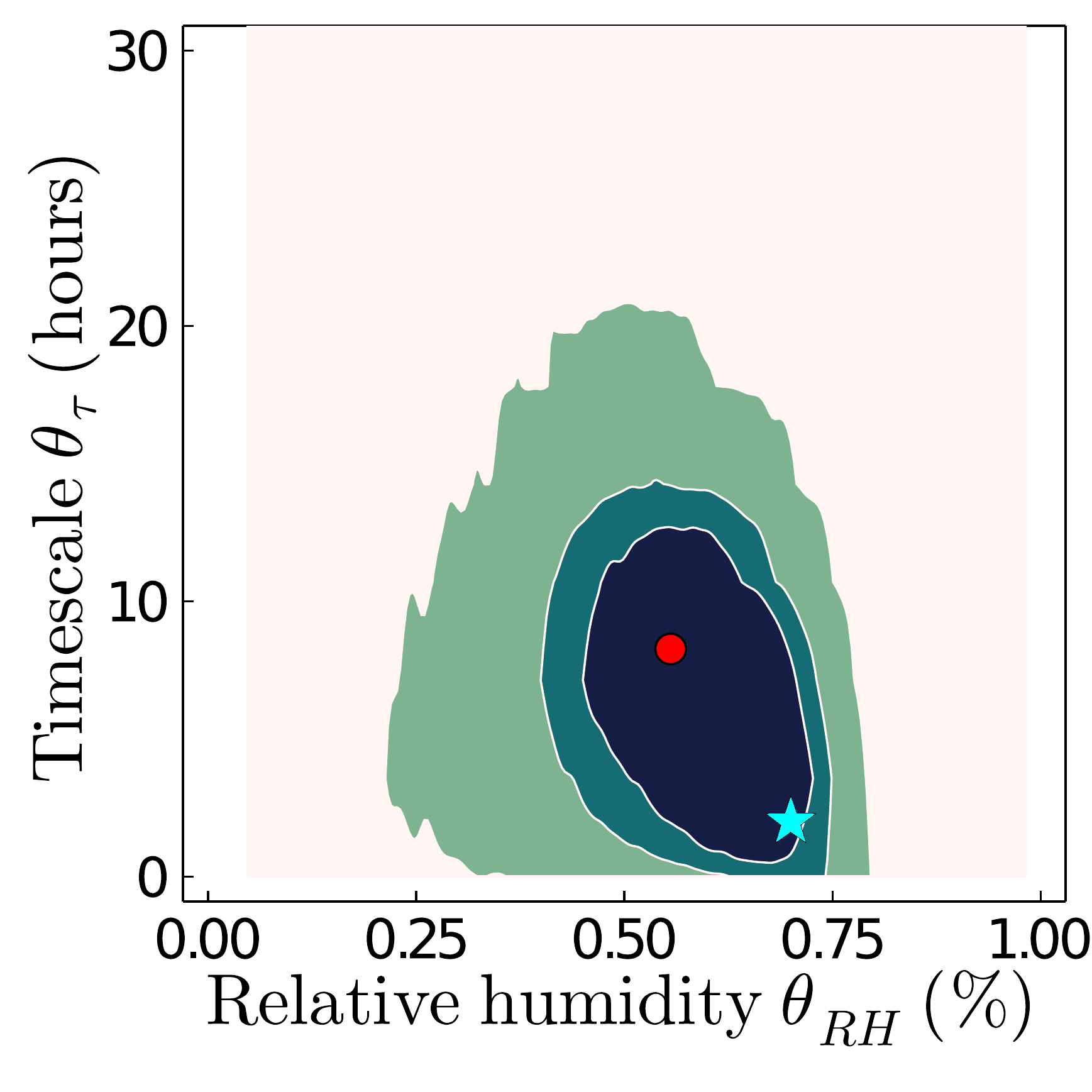} &
  \subfigimgtwo[width=\linewidth,valign=t]{(b)}{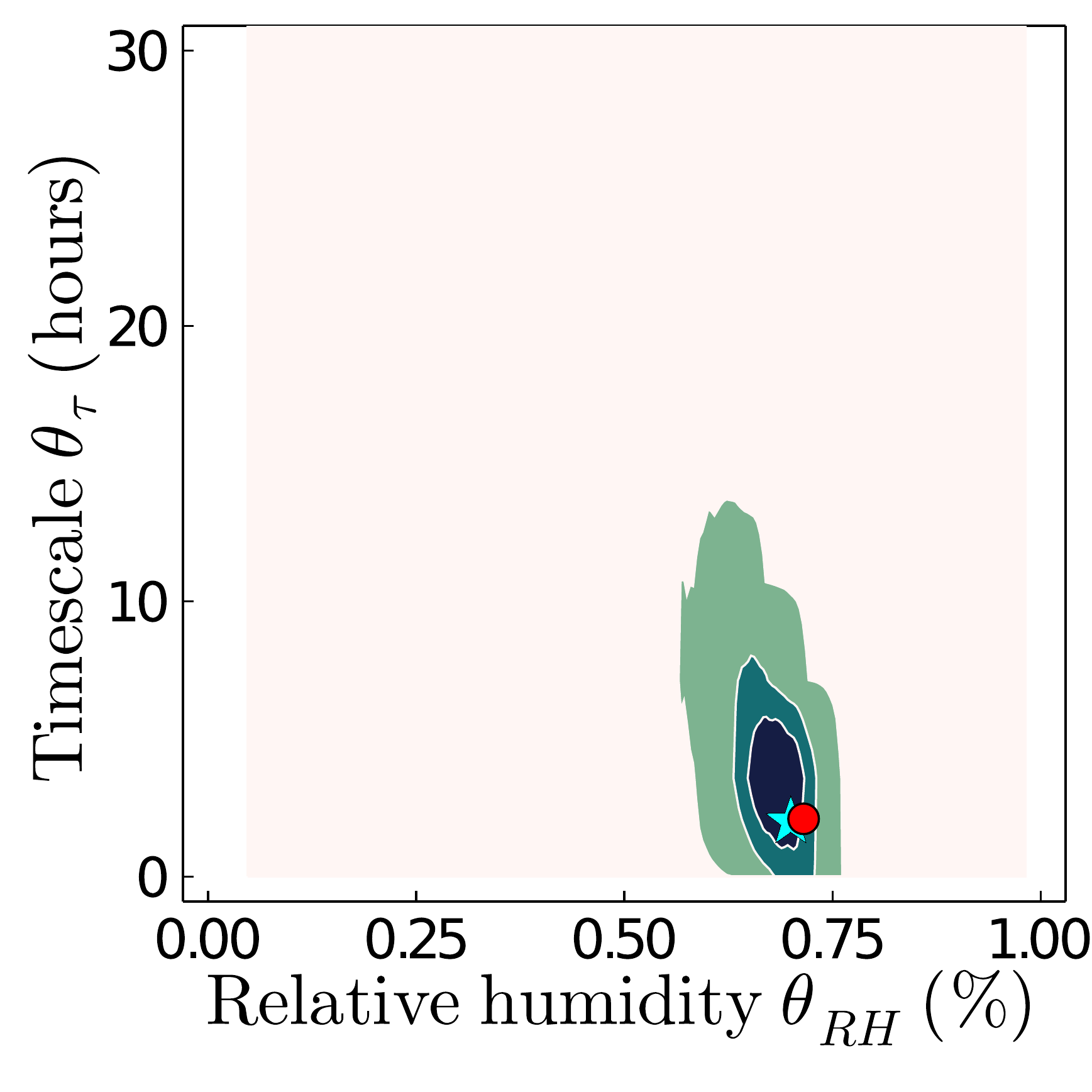} 
  \end{tabular}
  \caption{Posterior density for (a) $T=360~\mathrm{d}$ and (b) $T=90~\mathrm{d}$ for the less informative statistics.
  Contours correspond to $50\%$, $75\%$, and $99\%$ of the posterior distribution.
  The location of the true parameters are indicated by the star, and the circle is the average of the ensemble members after the last EKI iteration.
  }
    \label{fig:mcmc_wind_0p1}
\end{figure}

Posterior distributions computed from CES with GP training data generated by EKI for $T=360~\mathrm{d}$ and $T=90~\mathrm{d}$ are shown in Figure \ref{fig:mcmc_wind_0p1}.
We use SVD truncation for both $T=360~\mathrm{d}$ and $T=90~\mathrm{d}$ (\ref{sec:decorrelation}).
The posterior for $T=360~\mathrm{d}$ is significantly larger than for $T=90~\mathrm{d}$, indicating that the annually averaged precipitation and surface wind speed do not provide substantive information regarding the convective parameters.
As with EKI for $T=360~\mathrm{d}$, the posterior distribution mode has substantial error when compared to the true parameters.
For $T=90~\mathrm{d}$, the uncertainty for $\theta_{\mathrm{RH}}$ estimated with the less informative data has collapsed to a similar order as the uncertainty estimated with informative data.
While the uncertainty is larger for $\theta_\tau$ when estimated using the less informative data, the overall posterior area is reasonably collapsed, as it is of similar order to the $T=360~\mathrm{d}$ posterior with informative data.
These results suggest that seasonally averaged precipitation rate and surface wind speed are sufficient statistics to estimate convective parameters while annually averaged precipitation and wind speed introduce higher estimation error and parameter uncertainty.

As with the informative GCM statistics, we compare the sizes of the posterior distributions by computing the two-dimensional area of the convex hull containing $75\%$ of the posterior mass.
The posterior areas, normalized by the prior area, averaged over the four CES instantiations, are shown in Table \ref{tab:T_360_informative}.
The posterior area for $T=360~\mathrm{d}$ is approximately $5$ times larger than the area produced with $T=90~\mathrm{d}$.
The implications of the reduction in the size of the posterior distribution on the parametric uncertainty in the GCM are tested in section \ref{sec:prediction}.

\subsubsection{Prediction experiments}
\label{sec:prediction}

\begin{figure}
    \centering
    \includegraphics[width=0.5\linewidth]{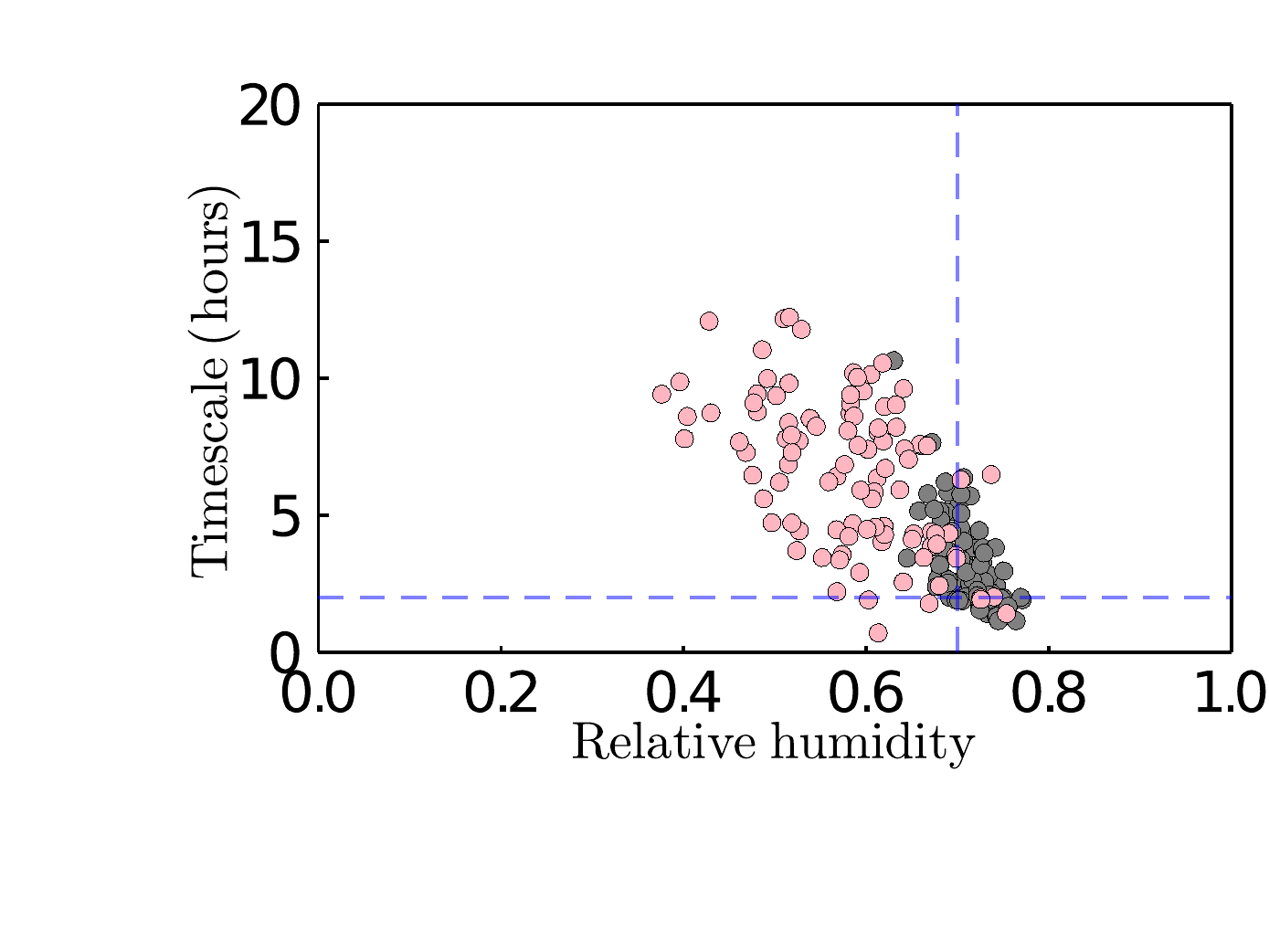}
    \caption{Samples of size 100 from the posterior distributions for
    prediction experiments.
    Black are samples from the $T=90~\mathrm{d}$ posterior and red are samples from the $T=360~\mathrm{d}$ posterior.
    The true parameters are shown with dashed blue lines.}
    \label{fig:prediction_params}
\end{figure}

To demonstrate the effect of parametric uncertainty in the GCM on climate predictions, we draw samples of $100$ parameter pairs from the posterior distributions resulting from the less informative statistics of precipitation rate and surface wind speed (Figure \ref{fig:mcmc_wind_0p1}). 
The samples of parameter pairs using seasonal integration ($T=90~\mathrm{d}$) and annual integration ($T=360~\mathrm{d}$) in the UQ are shown in Figure \ref{fig:prediction_params}.
With each sample of parameters, we produce climate quantities of interest, averaged over a $20$ year period.
We compute ensemble statistics over the outputs from the samples (prediction uncertainty) and compare these with a $20$ year simulation with the true parameters $\bm{\theta}^\dagger$.

\begin{figure}
  \centering
  \begin{tabular}{@{}p{0.33\linewidth}@{\quad}p{0.33\linewidth}@{\quad}p{0.33\linewidth}@{}}
  \subfigimgthree[width=\linewidth,valign=t]{(a)}{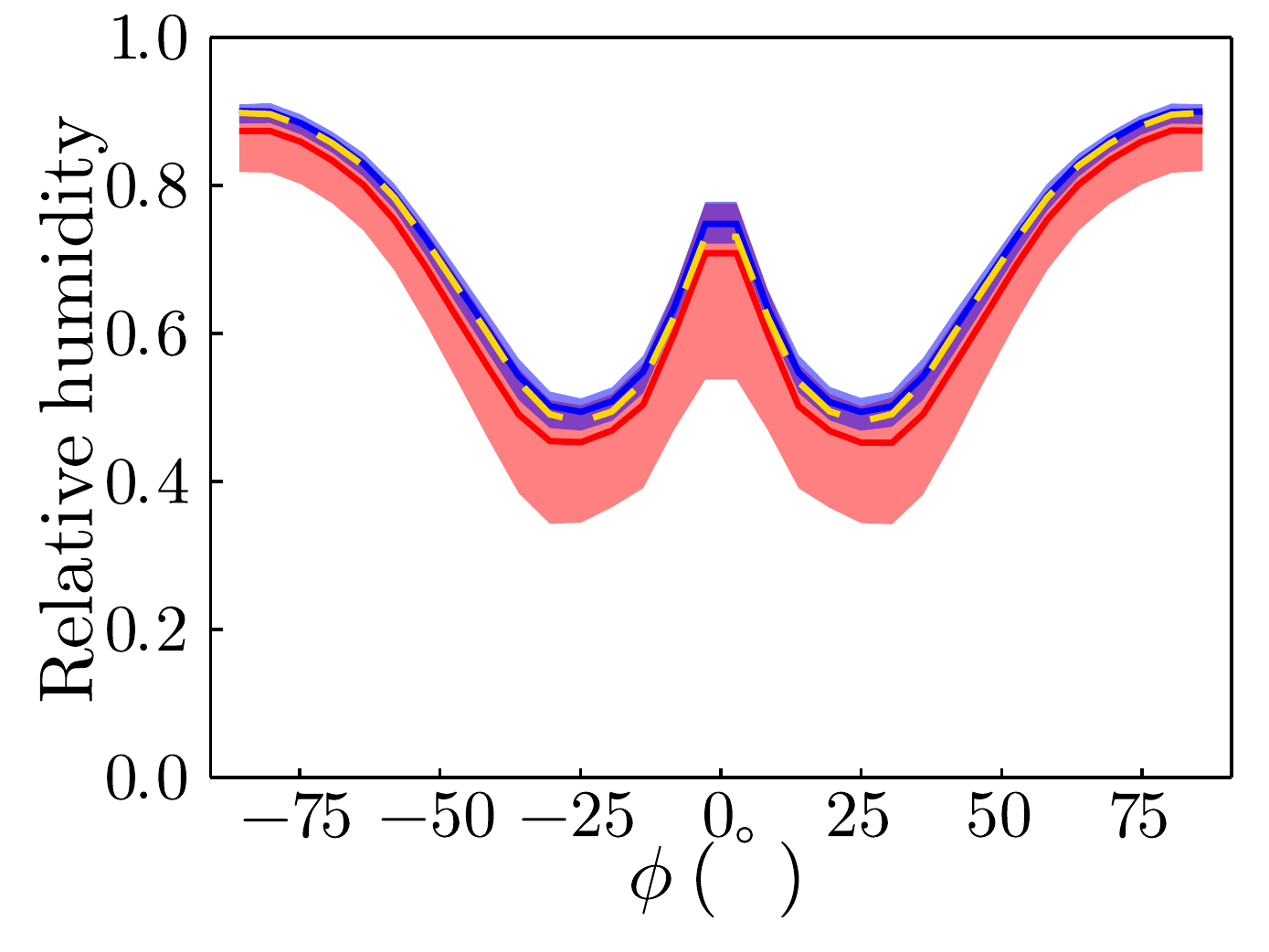} &
  \subfigimgthree[width=\linewidth,valign=t]{(b)}{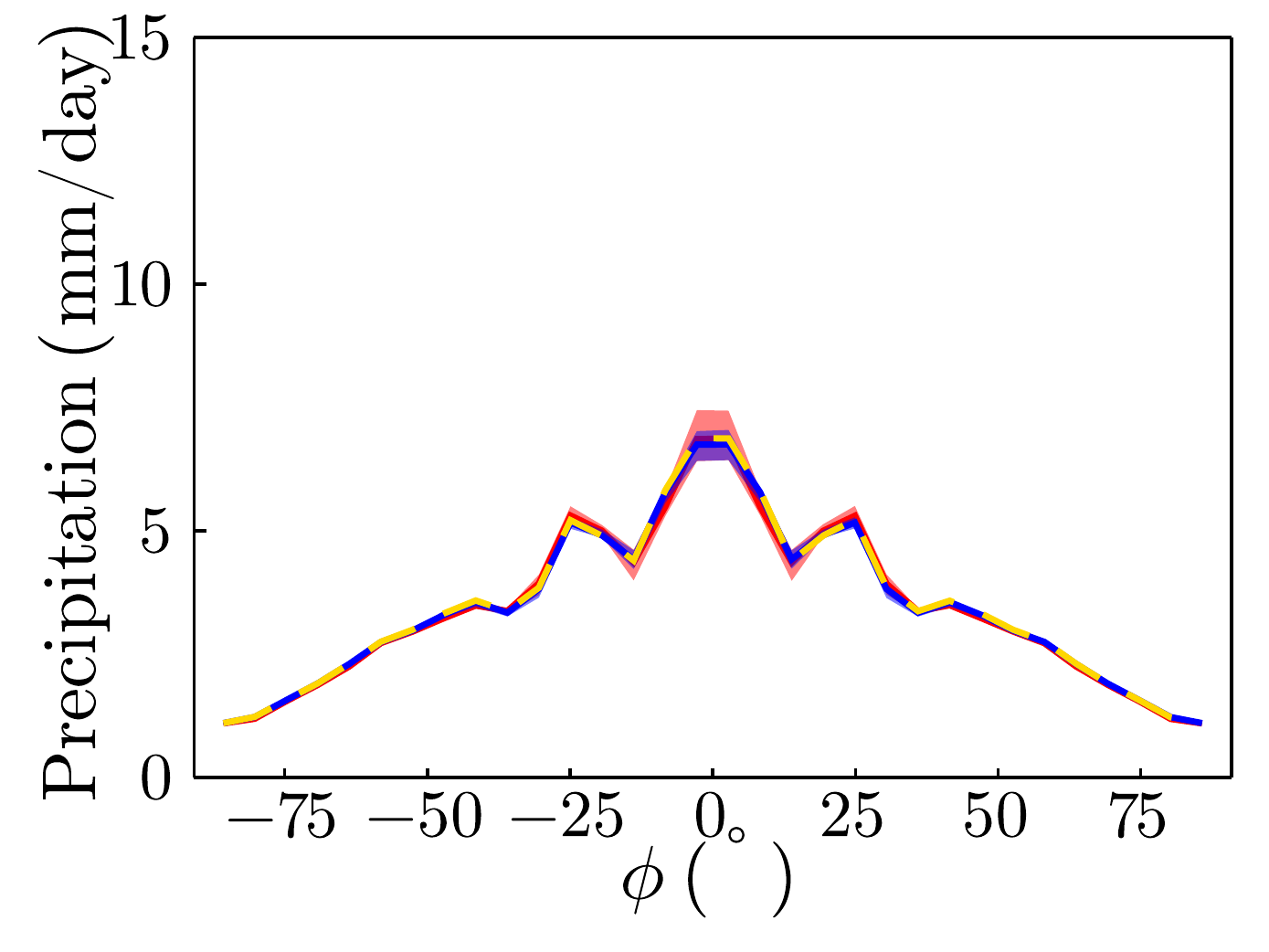} &
  \subfigimgthree[width=\linewidth,valign=t]{(c)}{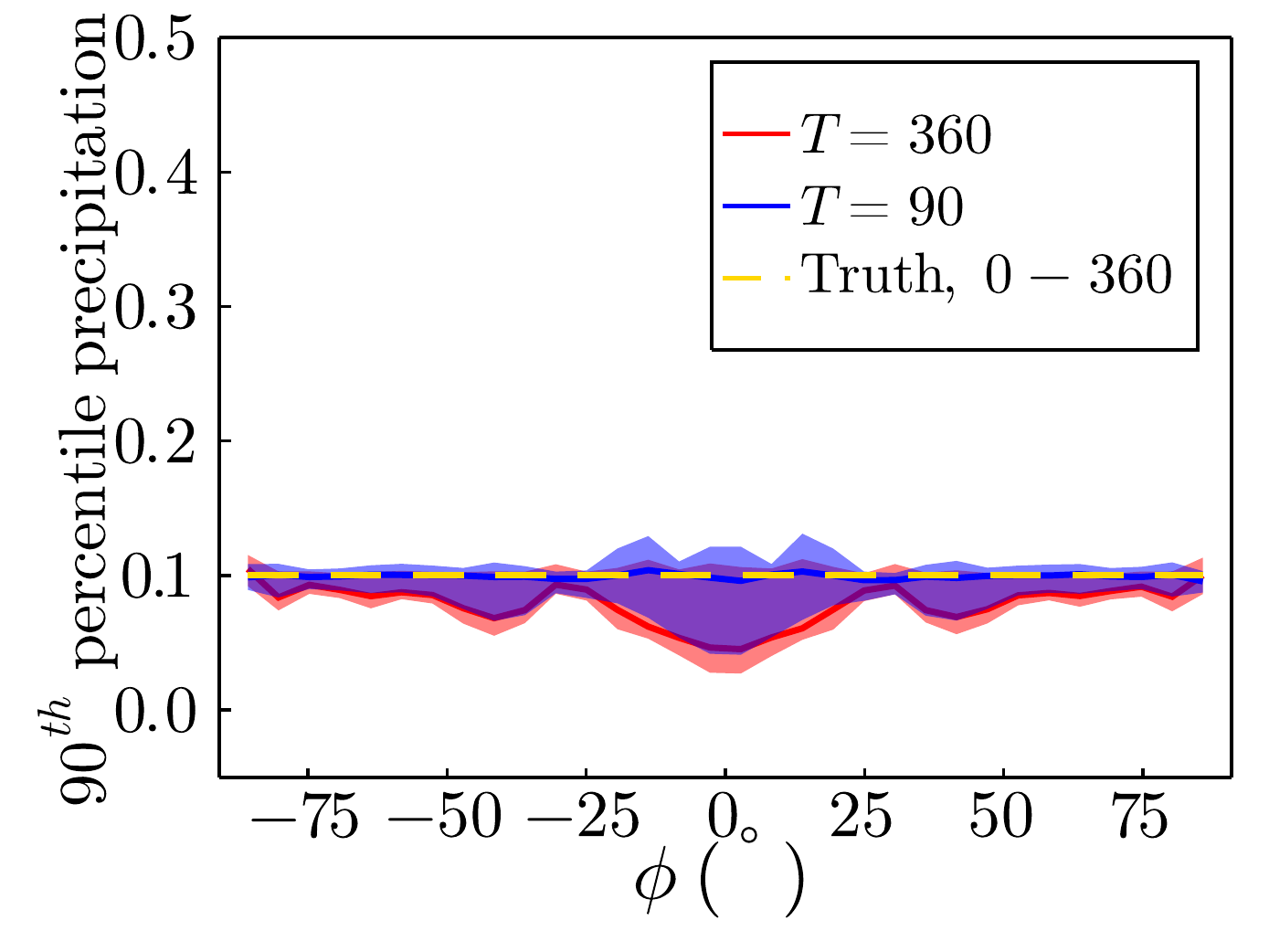}
  \end{tabular}
  \caption{Prediction experiments for the case with no imposed warming for (a) relative humidity at $\sigma_p=0.5$, (b) precipitation rate, and (c) intense precipitation probability.
  The climate statistics are averaged over $20$ years.
  The posteriors from the less informative statistics (surface wind speed and precipitation rate) are used.
  The lines correspond to the mean of the predictions and the shaded regions, correspond to $95\%$ confidence intervals.
  The climate statistics at the true parameters $\bm{\theta}^\dagger$ are shown with a dashed line.
  }
    \label{fig:predictions}
\end{figure}

The relative humidity at $\sigma_p=0.5$, the precipitation rate, and the intense precipitation probability (the probability of exceeding a 90th percentile threshold averaged over the $20$ year simulation at $\bm{\theta}^\dagger$) are shown in Figure \ref{fig:predictions}. 
The posterior distribution was estimated using (seasonally averaged) precipitation rate data directly, but neither relative humidity nor intense precipitation probability were used.
The mean absolute percent errors for the mean prediction values, averaged across latitudes, for the $T=90~\mathrm{d}$ posterior are $1.2\%$, $0.6\%$, and $1.4\%$, for the relative humidity, precipitation rate, and intense precipitation probability, respectively.
For the predictions with the $T=360~\mathrm{d}$ posterior, the mean absolute percent errors are $4.7\%$, $1.8\%$, and $21.5\%$.

The prediction uncertainty in each climate quantity of interest is indicated by a $95\%$ confidence interval, estimated as the values of the $2.5^{th}$ and $97.5^{th}$ percentiles of the variables of interest at each latitude over the convective parameter pairs (Figure \ref{fig:prediction_params}).
For both $T=90~\mathrm{d}$ and $T=360~\mathrm{d}$, there is limited prediction uncertainty in the mean precipitation rate.
The widths of the confidence intervals are $6\%$ and $4\%$ of the mean precipitation rate, averaged across latitudes, for $T=360~\mathrm{d}$ and $T=90~\mathrm{d}$, respectively.
The prediction uncertainties in relative humidity and intense precipitation probability are significantly higher for UQ performed with annually averaged data ($T=360~\mathrm{d}$) compared to UQ with seasonally averaged data ($T=90~\mathrm{d}$).
The $T=90~\mathrm{d}$ posterior reduces the size of the $95\%$ confidence interval for the relative humidity by $70\%$, averaged across latitudes, compared to $T=360~\mathrm{d}$.
The reduction in the size of the $95\%$ confidence interval for the intense precipitation probability is $9\%$.

We also performed idealized global-warming prediction experiments.
To represent global warming in the idealized GCM, the longwave opacity in the atmosphere is increased by $50\%$, as in \cite{o2008hydrological} which results in a global-mean surface air temperature increase from 284~K in the control climate to 292~K in the warm climate.
We accumulated GCM statistics over a $20$ year window after a spinup of one year, running simulations with the true GCM parameters $\bm{\theta}^\dagger$ with increased longwave opacity for comparison. We use the same 100 parameter pairs drawn from the posterior distributions estimated from the less informative climate statistics.

\begin{figure}
  \centering
  \begin{tabular}{@{}p{0.33\linewidth}@{\quad}p{0.33\linewidth}@{\quad}p{0.33\linewidth}@{}}
  \subfigimgthree[width=\linewidth,valign=t]{(a)}{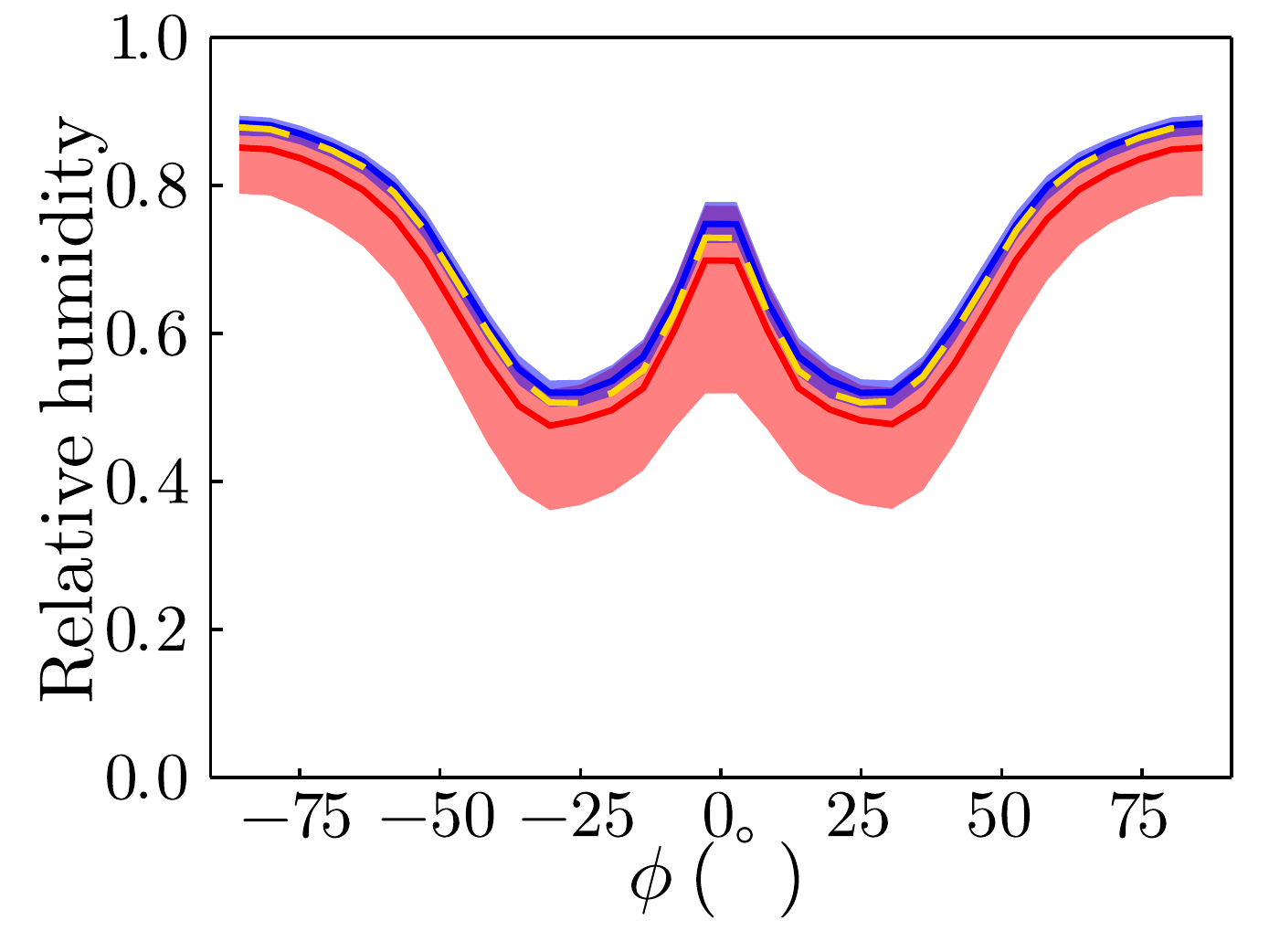} &
  \subfigimgthree[width=\linewidth,valign=t]{(b)}{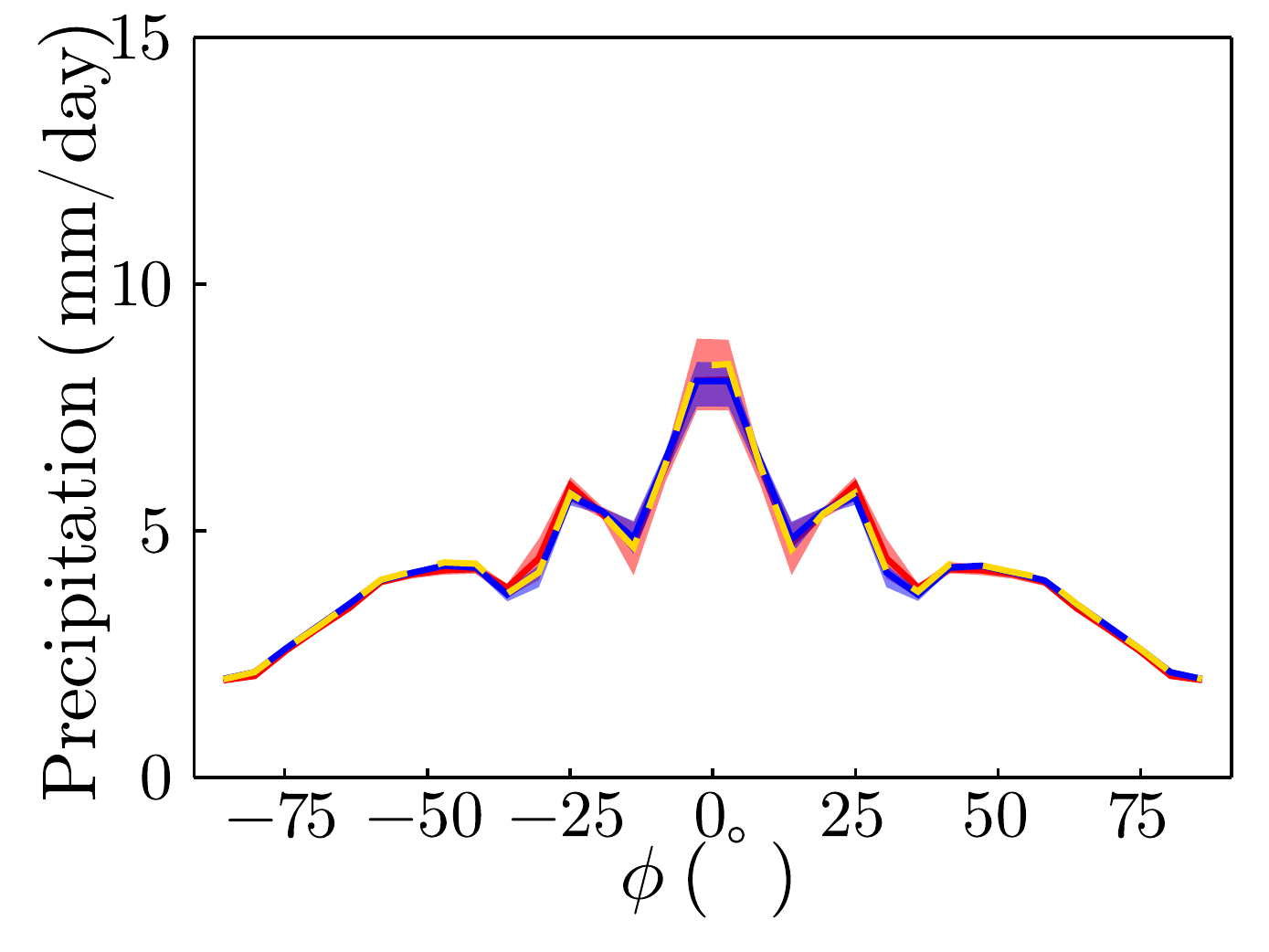} &
  \subfigimgthree[width=\linewidth,valign=t]{(c)}{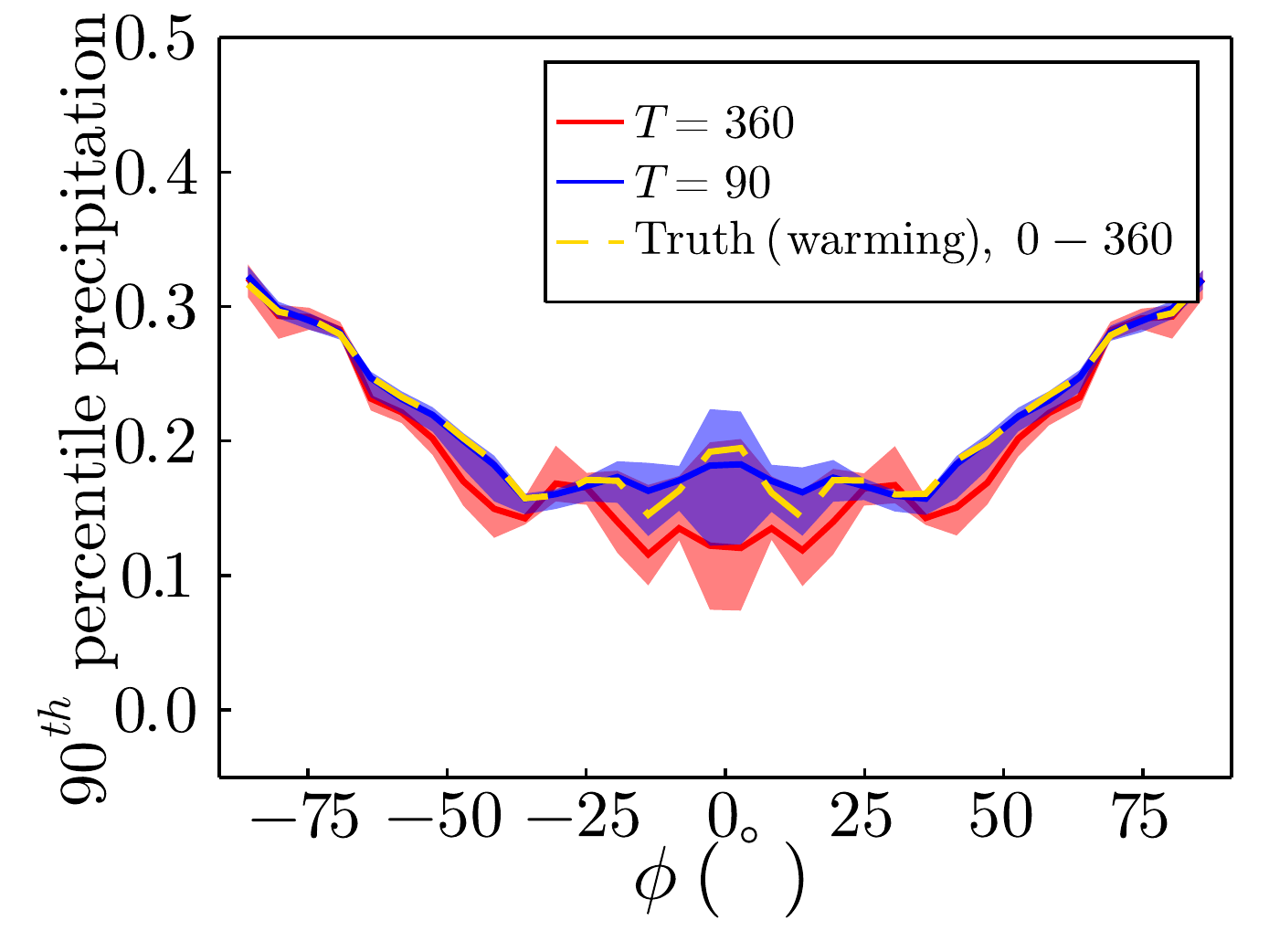}
  \end{tabular}
  \caption{
  Prediction experiments for the case with imposed warming ($50\%$ increase in longwave opacity) for (a) relative humidity at $\sigma_p=0.5$, (b) precipitation rate, and (c) intense precipitation probability.
  The climate statistics are averaged over $20$ years.
  The posteriors from the less informative statistics (surface wind speed and precipitation rate) are used.
  The lines correspond to the mean of the predictions, and the shaded regions correspond to $95\%$ confidence intervals.
  The climate statistics at the true parameters $\bm{\theta}^\dagger$ are shown with a dashed line.
  }
    \label{fig:predictions_warming}
\end{figure}

The idealized global warming results are shown in Figure \ref{fig:predictions_warming}.
As expected, the warming climate predictions with the $T=90~\mathrm{d}$ posterior distribution are more accurate than with the $T=360~\mathrm{d}$ posterior, especially for the relative humidity and intense precipitation probability.
The mean absolute percent error for the mean prediction values, averaged across latitudes, for the $T=90~\mathrm{d}$ posterior are $1.7\%$, $1.1\%$, and $2.2\%$, for the relative humidity, precipitation rate, and intense precipitation probability, respectively.
For the predictions with the $T=360~\mathrm{d}$ posterior, the mean absolute percent errors are $4.7\%$, $2.2\%$, and $10.4\%$.
The $T=90~\mathrm{d}$ posterior reduces the size of the $95\%$ confidence interval for the relative humidity in a warmed climate by $75\%$ compared to $T=360~\mathrm{d}$, averaged across latitudes.
The reduction in the size of the $95\%$ confidence interval for the intense precipitation probability in a warmed climate is $35\%$.
These results demonstrate that the incorporation of the seasonal cycle in the UQ reduces the parametric uncertainty in climate predictions in this model.

\section{Conclusions}
\label{sec:conclusions}

We performed calibration and UQ of a convective parameterization in a seasonally forced idealized GCM.
While GCMs are typically tuned using annually and globally averaged climate statistics \cite{hourdin2017art}, our results demonstrate, in an idealized setting, the qualitative, quantitative and systematic refinement of parameter distributions through the incorporation of seasonal information.
Performing parameter calibration with seasonally averaged data significantly reduced the error associated with the estimated parameters compared to calibration using annually averaged data.
The posterior distributions resulting from the Bayesian UQ with seasonally averaged data were reduced in size.
One measure of success for Bayesian UQ, is to capture the true parameters within a high mass region of the posterior distribution (demonstrated in our results). 
We are also interested in the shape and size of the posterior distribution, as this provides valuable information about parameter correlations and uncertainty, with respect to observed data. 
We demonstrate that choosing suitable data for UQ which reduce the size of the posterior distribution can lead to significant reductions of climate prediction uncertainty.

The impact of incorporating additional frequency content is pronounced when the climate statistics used for UQ are less informative about the parameterizations. 
Such situations often occur in the climate modeling setting where a number of parameterizations may simultaneously and nonlinearly influence a quantity of interest and where it is not always clear which climate statistics should be used to calibrate an unknown GCM parameter.

To enable UQ in the climate model setting, we used the calibrate-emulate-sample (CES) methodology \cite{cleary2021calibrate}, which enables the efficient and accurate estimation of Bayesian posterior distributions of parameters from noisy climate data.
CES uses gradient-free optimization to calibrate parameters and generate parameter-data pairs, Gaussian process regression to emulate the parameter-to-data mapping, and Markov Chain Monte Carlo (MCMC) to sample from the posterior distribution.
The emulation and sampling is performed in a decorrelated dataspace through a transformation based on the principal component analysis of the noise covariance matrix.
In this study, we modified the original CES methodology by first normalizing the data to ensure all statistics are weighted equally in the UQ and second by regularizing the covariance matrix to enable the UQ of applications with ill-conditioned or rank deficient covariance matrices.
We quantified the impact of the normalization and regularization, with the regularization smoothing the posterior and slightly increasing its size. 

Beyond convection, Earth system models rely on a number of parameterizations of cyclostationary processes, including models of carbon accumulation and storage \cite{bloom2016decadal} and of atmospheric boundary layer (ABL) turbulence which parameterize unclosed subgrid processes \cite{stull2012introduction}.
Carbon storage is inherently seasonal \cite{rowland2014evidence}.
The ABL has distinct seasonal and diurnal variations \cite{wyngaard2010turbulence, howland2020influencefield}. 
We expect our results to be relevant to calibration and UQ for statistical variations on different timescales, beyond the seasonal cycle.
For example, based on our findings presented here, we anticipate that incorporating the diurnal variation of ABL turbulence statistics may improve the UQ of subgrid scale turbulence models in GCMs.
More broadly, we anticipate the findings of this study to be relevant to UQ in problems with multi-scale temporal dynamics.

The selection of the optimal aggregation timescale for climate statistics and choice of objective function remain open questions.
In the context of parameter estimation, the integration timescale used to generate time-averaged statistics is a hyperparameter that dictates the trade-off between the frequency content and signal-to-noise ratio of the climate statistics.
In the limiting case of $T\rightarrow 0$, the influence of the initial conditions on the state is pronounced, as in weather forecasting \cite{houtekamer2016review}.
Conversely, as $T\rightarrow \infty$, all frequency content is removed from the statistics, which, as this study demonstrates, can adversely impact parameter estimation.
We anticipate that the selection of the filter timescale $T$ may be problem and parameterization specific, as it relates to the question of parameter identifiability.
In this study, we selected seasonal averages because of the large amplitude of the seasonal cycle in many climate statistics and indications that seasonal variations are informative about the climate change response in many climate variables \cite{Schneider21b}.
Future work should develop a more generalized approach for the selection of $T$ based on the frequency content of the quantities of interest.

\FloatBarrier

\appendix

\section{Data and model output decorrelation for rank deficient problems}
\label{sec:decorrelation}

To facilitate the transformation into the uncorrelated space with rank deficient covariance matrices, the singular value decomposition/principal component analysis is truncated.
Since the statistical quantities of interest have a range of magnitudes, the data are first normalized, and then the singular value decomposition is truncated as a form of regularization \cite{hansen1987truncatedsvd}.
A detailed investigation of the influence of the normalization and truncation on the posteriors resulting from CES is performed in \ref{sec:trunc_vs_reg}.


The data used for UQ are provided concurrently to the CES pipeline in the concatenated vector $\bm{y}$.
In this framework, the statistical quantities in the data $\bm{y}$ may have a range of magnitudes, and normalization is required before regularization (e.g., \cite{tibshirani1996regression}).
The data are normalized by a characteristic value associated with each individual data type $\bm{y}_c \in \mathbb{R}^{N}$.
The normalized data are
\begin{align}
{y}^*_i = {y}_i \cdot {y}_{c,i}^{-1},
\label{eq:normalization}
\end{align}
where $(^*)$ denotes normalized data.
The characteristic values in $\bm{y}_c$ are described in the application of CES to the seasonally forced GCM in section \ref{sec:gcm}.
Each column of $Y$ is normalized using Eq. \eqref{eq:normalization} to yield the normalized data matrix ${Y^*}$.
The normalized covariance is $\Sigma^* = \mathrm{cov}({Y^*})$.
The SVD transformation is performed using covariance matrix $\Sigma^*$ and data $\bm{y}^*$. 


The SVD is truncated in order to account for rank deficient or ill-conditioned covariance matrices \cite{hansen1987truncatedsvd}.
The truncated SVD is defined for covariance $\Sigma^*$
\begin{align*}
\Sigma^*  &\approx \Sigma_k^* = V_k D_k^{*2} V_k^\intercal &
D_k^* &= \mathrm{diag}(\sigma_1^*, ..., \sigma_k^*),
\end{align*}
truncated at the $k\leq \mathrm{rank}(\Sigma^* )$ singular value.
The truncated singular vector matrix is $V_k = \left[ v_1, ..., v_k \right]$ where $\bm{v}_i$ is the singular vector corresponding to the singular value $\sigma_i^{*2}$.
The truncated SVD space is given by
\begin{align}
\tilde{{\Sigma}}_k &= {D}_k^{-1} V_k^\intercal {\Sigma^*} V_k {D}_k^{-1} \\
\tilde{{\bm{y}}}_k &= {D}_k^{-1} V_k^\intercal {\bm{y}^*}.
\end{align}
In this study, the truncation location is selected as the lowest value of $k$ such that $\sum_{i=1}^{k} \sigma_i^{*2} \geq 0.95 \sum_{i=1}^{N} \sigma_i^{*2}$, so that $95\%$ of the variance is retained.
The GP output can be mapped to the original normalized space using
\begin{align}
{\mathcal{G}}_{\mathrm{GP}}^* &= V_k {D}_k \tilde{\mathcal{G}}_{\mathrm{GP}} \\
{\Sigma}_{\mathrm{GP}}^* &= V_k {D}_k  \tilde{{\Sigma}}_{\mathrm{GP}} {D}_k V_k^\intercal.
\end{align}
Finally, the GP output can be transformed into the dimensional space through $\mathcal{G}_{\mathrm{GP}} = {\mathcal{G}}_{\mathrm{GP}}^* \odot \bm{y}_c$ and $\Sigma_{\mathrm{GP}} = {\Sigma}_{\mathrm{GP}}^* \odot \bm{y}_c \bm{y}_c^\intercal$, where $\odot$ denotes pointwise multiplication.

\section{Sensitivity of the posterior to covariance regularization}
\label{sec:trunc_vs_reg}

\begin{figure}
    \centering
    \includegraphics[width=0.5\linewidth]{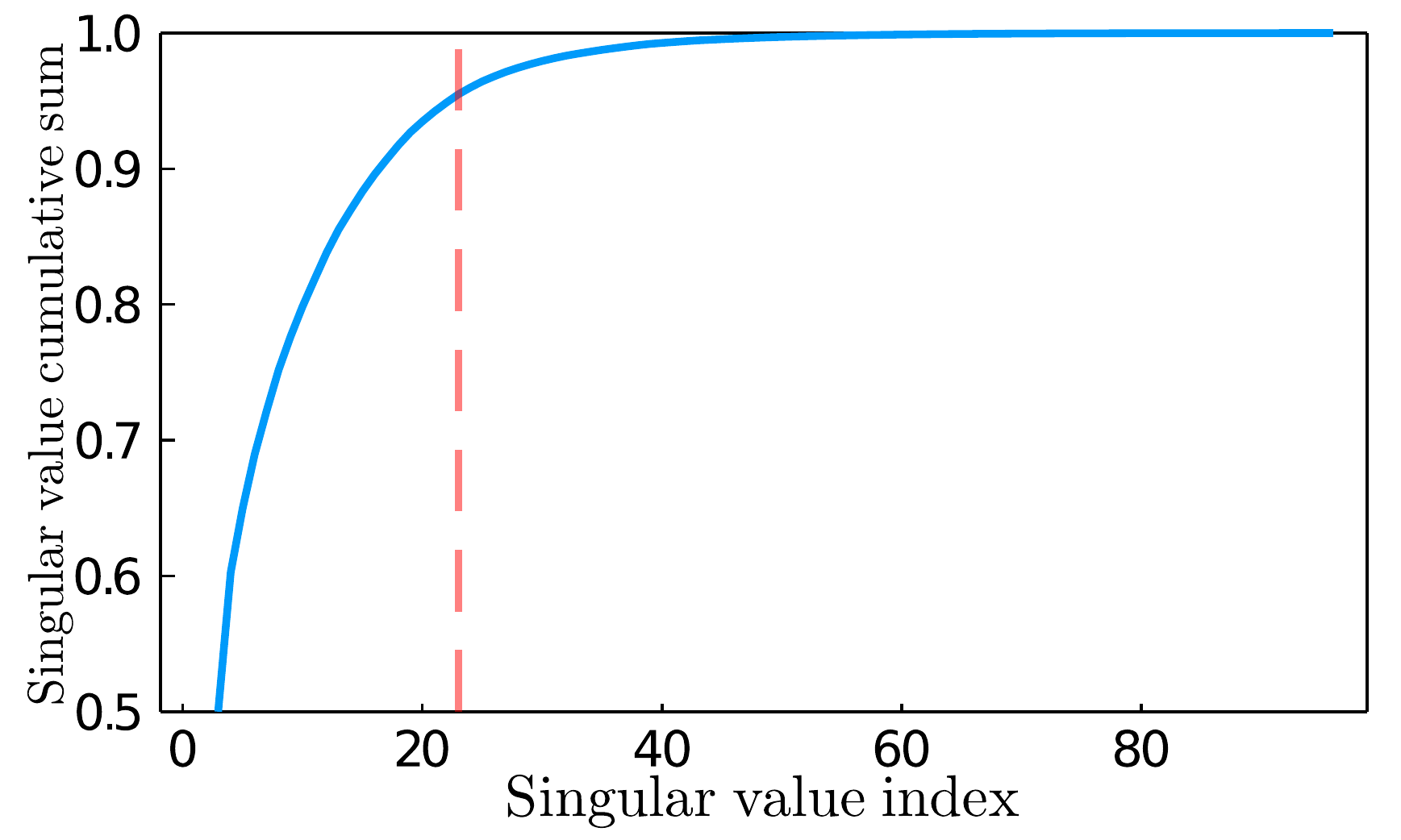}
    \caption{Cumulative sum of the singular values from the normalized $T=360~\mathrm{d}$ synthetic data.}
    \label{fig:singular_values_normalized}
\end{figure}

\begin{figure}
    \centering
    \includegraphics[width=0.66\linewidth]{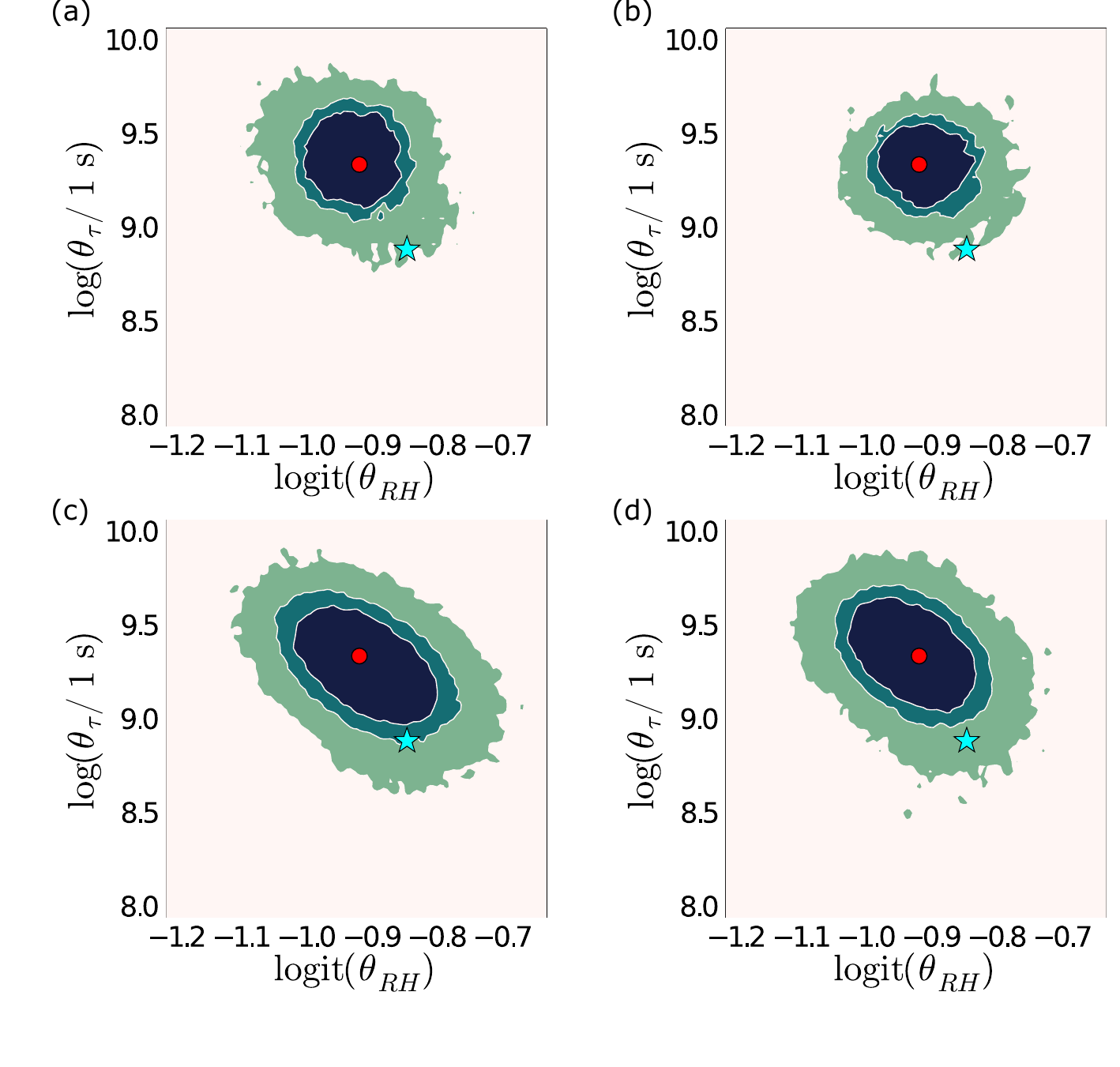}
    \caption{Posterior distributions using the informative statistics with $T=360~\mathrm{d}$. 
  Comparison of the posteriors estimated with (a) original, (b) normalized, (c) normalized and regularized (truncation), and (d) normalized and regularized (Tikhonov diagonal inflation) covariance matrix $\Sigma$.
  }
    \label{fig:base_vs_trunc}
\end{figure}

In section \ref{sec:emulate}, the SVD is truncated as a form of regularization.
Here, we detail the sensitivity of the CES posterior distributions to the SVD regularization using informative GCM statistics (see Table \ref{tab:statistics}) and annually averaged data with $T=360~\mathrm{d}$, for which the data covariance matrix is full rank.
The data are normalized, as discussed in \ref{sec:decorrelation}.
The cumulative sum of the singular values is shown in Figure \ref{fig:singular_values_normalized}.
Truncation at $95\%$ of the singular value energy corresponds to $k=23$.
Regularization can also be performed using Tikhonov regularization, which inflates the diagonal elements \cite{hansen1987truncatedsvd,Schneider01b}.
Following \cite{hansen1987truncatedsvd}, regularization in the form of diagonal inflation ${\Lambda} = \mathrm{diag}(\lambda^2)$ is added to ${\Sigma}$ where
$\lambda = (\sigma_k^3 \sigma_{k+1})^{1/4}.$
The SVD is performed such that
\begin{equation}
{\Sigma} + {\Lambda} = V {D}^2 V^\intercal,
\end{equation}
where $D^2$ is a diagonal matrix of the singular values $\sigma_i^2 + \lambda^2$.
The orthonormal eigenvectors are in $V$.
The influence of the truncation and Tikhonov regularizations on the posterior areas for $T=360~\mathrm{d}$ are provided in Table \ref{tab:T_360_informative_reg}.
The $T=360~\mathrm{d}$ posterior distributions are shown in Figure \ref{fig:base_vs_trunc}.
The regularization slightly increases the posterior area and smooths the posterior roughness.

\begin{table}
  \begin{center}
\def~{\hphantom{0}}
  \begin{tabular}{c|cccc}
  \toprule
    Parameter space & Original & Normalized & Truncated & Tikhonov \\
    \hline
    Informative data & $0.5\%$ & $0.5\%$ & $0.8\%$ & $0.9\%$ \\
    \toprule
  \end{tabular}
  \caption{
  The ratio ($\%$) of the area occupied by the posterior to the area occupied by the prior.
  The area of the convex hull containing $75\%$ of the posterior mass for each regularization method is normalized by the area containing $75\%$ of the prior mass.
  }
  \label{tab:T_360_informative_reg}
  \end{center}
\end{table}

\section{Parameter estimation with filtering and smoothing}
\label{sec:filtering}

It is often of practical interest to estimate parameters in problems that are not statistically stationary, such as the seasonally forced GCM with time-dependent boundary conditions.
In problems which are not statistically stationary (e.g., with time-evolving boundary conditions), all requisite data for parameter UQ may not be available at the initialization of the estimation.
This arises in the EnKF setting for state estimation (e.g., \cite{houtekamer2016review}).

We can instead pose EKI as a filtering approach for parameter estimation where the synthetic data are collected sequentially in a time-dependent problem.
This differs from the definition of the data in section \ref{sec:seasonal_inverse}, where the data from each season are collected together into $\bm{y}$ (termed smoothing).
For the filtering approach, the data is observed and averaged over time $t_j\rightarrow t_j+T$ in $\bm{y}_{T,j}$, where $j$ indicates the time of the year in the GCM simulation.
The EKI update is performed using $\bm{y}_{T,j}$ and $\mathcal{G}_{T,j}(\bm{\theta}^{(k)})$ to update $\bm{\theta}^{(k+1)}$, and this process is repeated for the specified number of EKI iterations.
The smoothing approach leverages the data averaged in each season concurrently ($\bm{y}$ and $\mathcal{G}(\bm{\theta}^{(k)})$) whereas the filtering approach uses the data averaged in each season sequentially ($\bm{y}_{T,j}$ and $\mathcal{G}_{T,j}(\bm{\theta}^{(k)})$), as it becomes available.

\begin{figure}
  \centering
  \begin{tabular}{@{}p{0.5\linewidth}@{\quad}p{0.5\linewidth}@{}}
  \subfigimgtwo[width=\linewidth,valign=t]{(a)}{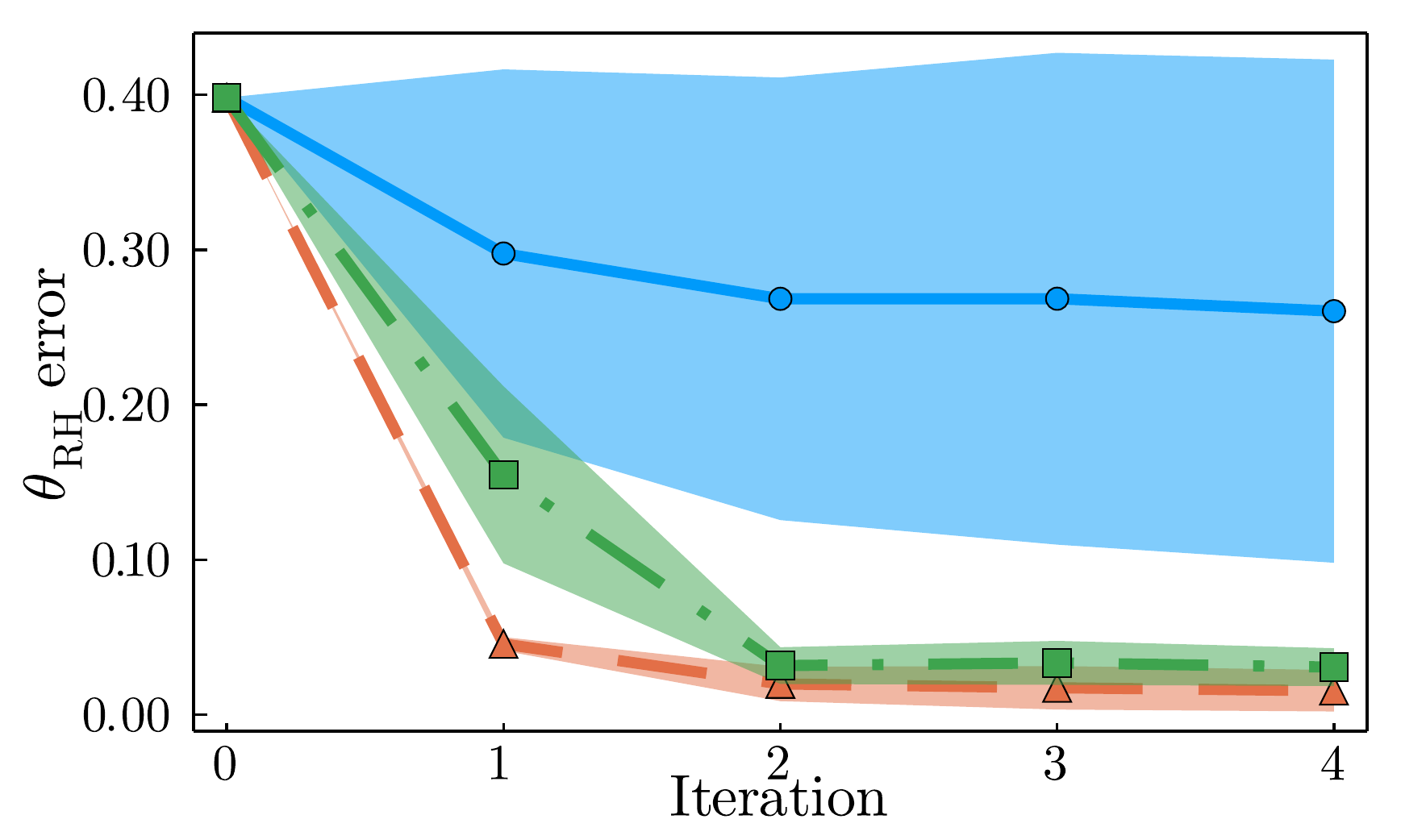} &
  \subfigimgtwo[width=\linewidth,valign=t]{(b)}{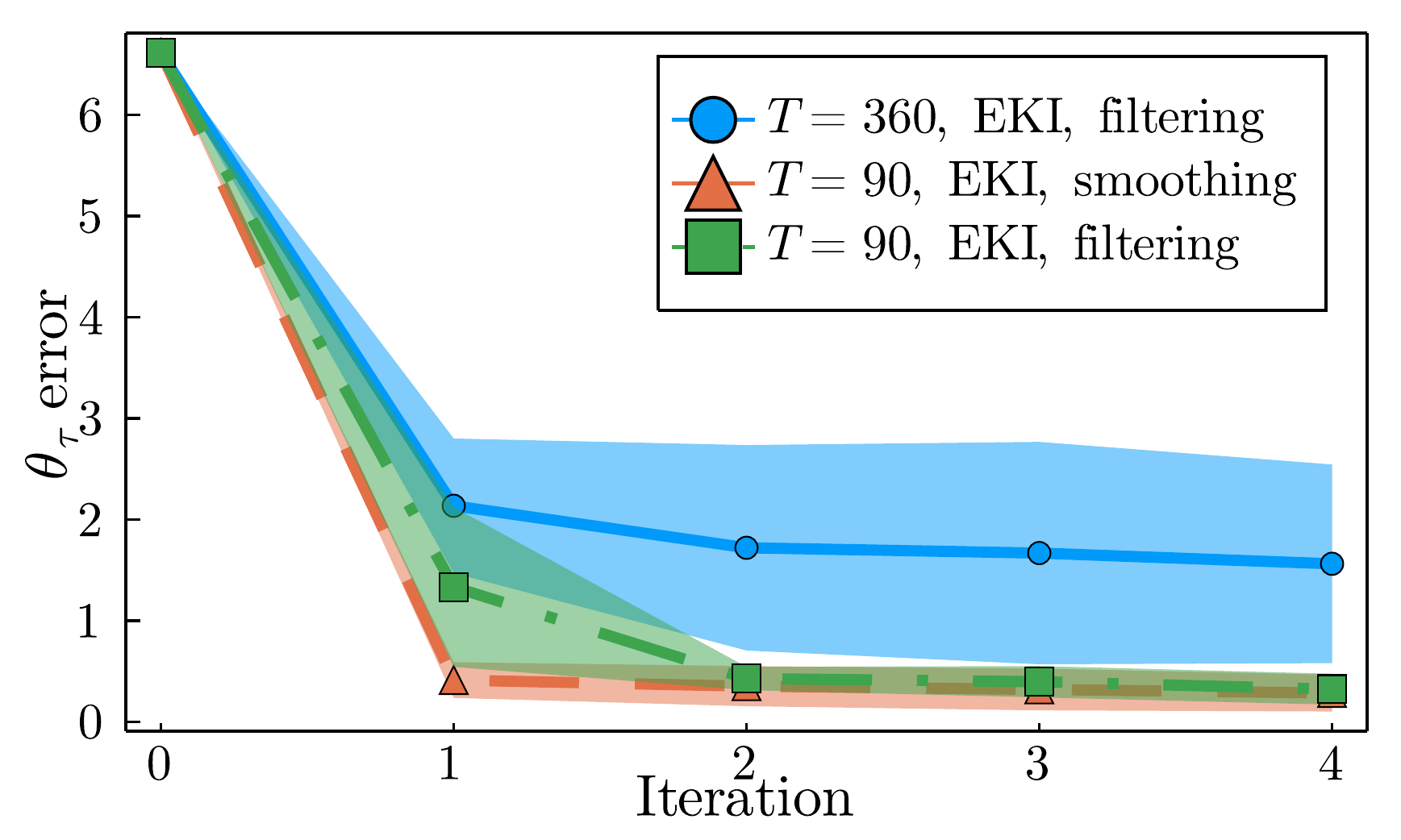}
  \end{tabular}
  \caption{Parameter calibration performed with EKI using less informative statistics.
  Mean square error of (a) the estimated relative humidity convective parameter $\theta_{\mathrm{RH}}$ compared to truth value $\theta_{\mathrm{RH}}^\dagger$ and of (b) the estimated relaxation time scale convective parameter $\theta_{\tau}$ compared to truth value $\theta_{\tau}^\dagger$.
  Solid lines are $T=360~\mathrm{d}$ and dashed lines are $T=90~\mathrm{d}$.
  }
    \label{fig:filtering}
\end{figure}

The parameter calibration results comparing filtering and smoothing are shown in Figure \ref{fig:filtering}.
For $T=360~\mathrm{d}$, filtering and smoothing are identical.
For $T=90~\mathrm{d}$, the filtering approach leverages the data from one season at a time to update the parameters $\bm{\theta}$.
The final parameter errors for the $T=90~\mathrm{d}$ filtering and smoothing approaches are similar, and are both significantly less than the error for the $T=360~\mathrm{d}$ case.
This result also demonstrates that the improvement in parameter estimates with $T=90~\mathrm{d}$, compared to $T=360~\mathrm{d}$, is a result of increased temporal information, rather than the dimensionality of the data space alone.
Depending on the application, the filtering approach may be beneficial for computational efficiency, since it reduces the length of the forward model simulations, instead of requiring temporal integration over the timescale corresponding to an impactful low-frequency cycle.

\bibliographystyle{unsrt}  
\bibliography{main}


\end{document}